\documentclass[11pt]{article}
\usepackage[body={16.6cm,23cm}]{geometry}

\usepackage[hang,small,center]{caption}
\usepackage{amsmath,amssymb}
\usepackage{amsfonts}
\usepackage{mathrsfs}
\usepackage{dsfont}

\usepackage{epic}
\usepackage{eepic}
\usepackage{graphicx}

\usepackage{pgf,pgfarrows}

\usepackage{cite}
\usepackage{enumerate}
\usepackage[curve]{xy}
\newcommand{\ds}{\displaystyle}

\renewcommand{\author}[1]{\large\rm #1\\ \bigskip}
\newcommand{\address}[1]{{\normalsize\it #1\\}\bigskip}
\renewcommand{\title}[1]{\bigskip\bigskip\Large\bf #1\bigskip\bigskip\\}

\newcommand{\Bigpsi}[3]{\phantom{\Psi}_2 \kern -.05em
\Psi_2\left(\genfrac{}{}{0pt}{}{#1}{#2}\biggl|#3\right)}

\newcommand{\bea}{\begin{eqnarray}}
\newcommand{\eea}{\end{eqnarray}}

\newcommand{\beq}{\begin{equation}}
\newcommand{\eeq}{\end{equation}}

\newcommand{\ii}{\mathsf{i}}

\newcommand{\hf}{\frac{1}{2}}

\newcommand{\qq}{{\mathsf q}}
\newcommand{\pp}{{\mathsf p}}

\newcommand{\iW}{\mathbb{W}}
\newcommand{\iS}{\mathbb{S}}
\newcommand{\iR}{\mathbb{R}}

\def\EXP{\textrm{{\large e}}}

\newcommand{\url}[1]{}

\renewcommand{\textcolor}[1]{}

\newcounter{app}
\newcounter{sapp}[app]
\def\theapp{\Alph{app}}
\newcommand{\app}[1]{
\refstepcounter{app}{\vspace{7mm}
\noindent\Large\bf Appendix
\theapp.
 \ #1 \par \vspace{5mm}}
\setcounter{equation}{0}
\def\theequation{\Alph{app}.\arabic{equation}}}

\begin{document}

\vglue 2cm

\begin{center}

\title{Elliptic gamma-function and 
multi-spin solutions of the Yang-Baxter equation}

\author {Vladimir V. Bazhanov}
\address{Department of Theoretical Physics,\\
         Research School of Physics and Engineering,\\
    Australian National University, Canberra, ACT 0200, Australia.\\
and\\
Mathematical Sciences Institute,\\
      Australian National University, Canberra, ACT 0200, Australia\\
E-mail: Vladimir.Bazhanov@anu.edu.au}
\author{Sergey M. Sergeev}
\address{Faculty of Information Sciences and Engineering,\\
University of Canberra, Bruce ACT 2601, Australia.\\
E-mail: Sergey.Sergeev@canberra.edu.au}

\end{center}

\begin{abstract}
We present a generalization of the master
solution to the quantum Yang-Baxter equation (obtained recently in 
arXiv:1006.0651) to the case of multi-component continuous spin
variables taking values on a circle.
The Boltzmann weights are expressed in terms of the elliptic gamma-function.
The associated solvable lattice model admits various equivalent
descriptions, including an interaction-round-a-face formulation with positive
Boltzmann weights. In the quasi-classical limit the model leads to a new series 
of classical discrete integrable equations on planar graphs.
\end{abstract}




\newpage
\section{Introduction}
There are solvable lattice models of statistical mechanics with only a
pair interaction between 
neighbouring spins, i.e, where two spins interact
only if they are connected by an edge of the lattice.  
The Yang-Baxter equation for these models usually 
takes the form of the ``star-triangle relation'' 
\cite{O44}.
The most notable discrete-spin models
in this class include the Kashiwara-Miwa \cite{Kashiwara:1986}
and chiral Potts \cite{vG85,AuY87,Baxter:1987eq} models (both of
them also contain the Ising model \cite{O44} and Fateev-Zamolodchikov
$Z_N$-model \cite{FZ82} as particular cases) see \cite{Bax02rip} for
a review.  
There are also important continuous spin models, including
Zamolodchikov's ``fishing-net'' model \cite{Zam-fish}, which describes
certain planar Feynman diagrams in quantum field theory, and the
Faddeev-Volkov model \cite{FV95}, connected with quantization
\cite{BMS07a} of discrete conformal transformations \cite{BSp,
  Steph:2005}. 

Recently \cite{Bazhanov:2010kz} we have found a new solution of the
star-triangle relation which contains as special cases all the discrete-
and continuous-spin solutions mentioned above\footnote{To be more
  precise, it only contains the solutions, which have a single
  one-dimensional spin at each lattice site. For this reason, it
  cannot contain the $D\ge2$ fishing-net model which has
  multi-dimensional spins.}, and also leads to new ones. This ``master
solution'' is expressed through elliptic gamma-functions and contains
two temperature-like variables.
It defines an exactly solvable lattice models with continuous spin variables
taking values on a circle. Its connection to
the theory of elliptic hypergeometric functions is discussed in
\cite{Spiridonov:2010em}.
From an algebraic point of view the model
is related 
to the modular double \cite{Faddeev:1999,Spiridonov-essays}
of the Sklyanin algebra \cite{Skl82}. The latter is an elliptic deformation of
the quantum group $U_q(sl(2))$, connected with the $R$-matrix of the
eight-vertex model \cite{Bax72}. 

In this paper we extend the main results of \cite{Bazhanov:2010kz} to
the case related to the $sl(n)$ algebra with $n\ge 3$. 
The generalized model has multi-component continuous spin
variables taking values on a circle. 
Similarly to the $n=2$ case of \cite{Bazhanov:2010kz} the model
contains two temperature-like parameters. 
The Boltzmann weights 
satisfy the so-called star-star relation
\cite{Baxter:1986phd,Bazhanov:1992jqa} (see Eq.\eqref{star-star}
below), which ensures the integrability
of the model. 
Currently, we claim this relation as a conjecture,
however, we expect that a complete proof could be obtained by a rather
straightforward generalization of the results of
\cite{Bazhanov:1990qk,Date:1990bs,Kashaev:1990wy,Bazhanov:1992jqa,Znbrok} 
devoted to discrete-spin models connected with the $sl(n)$ algebra.

It should be noted, that apart from the $n=2$ case previously
considered in \cite{Bazhanov:2010kz}, 
the two-spin Boltzmann weights are not real and positive. 
Fortunately, this is not an indication that the model is unphysical. 
It can be reformulated as an 
``interaction-round-a-face'' (IRF) model and then there exist
a domain of parameters, where the IRF-type 
Boltzmann weights become real and positive, see Section~\ref{sec:ybe}.

The quasi-classical (or low-temperature) limit of the model is
considered in Section~\ref{sec:quasi-classical}. A stationary spin 
configuration which gives the
leading contribution to the partition function in this limit is
described by new classical discrete integrable equations for
multi-component fields assigned to lattice sites. These equations
can be thought as a generalization of the Laplace-type equation associated
with the famous $Q_4$ system \cite{AdlerBobenkoSuris,AS04} to the
multi-component case. 

In Conclusion we summarize the results and discuss their connections
with some other integrable systems in two and three dimensions. 

\section{Formulation of the model}
We start with the definition of the elliptic gamma-function 
\cite{Ruijsenaars-elliptic,Felder-Varchenko,Spiridonov-beta,Spiridonov-essays}.
Let $\qq,\pp$ be two elliptic nomes (they play the role of 
the temperature-like parameters),
\begin{equation}
\pp\;=\;\EXP^{\ii\pi\sigma}\;,\quad
\qq=\EXP^{\ii\pi\tau}\;,\quad \mbox{Im}\,\sigma>0,  \quad \mbox{Im}\,\tau>0\,. 
\end{equation}
In principle, these parameters can be arbitrary (apart from the
requirements $|\pp|<1$ and $|\qq|<1$), however, in the following we
will often refer to special regimes  
when  $\pp$ and $\qq$ are either real or complex
conjugate to each other,
\beq
(i) \quad \pp^*=\pp,\quad \qq^*=\qq,\qquad (ii) \quad
\pp^*=\qq\,.\label{regimes}
\eeq
Note that in both of these cases the ``crossing parameter'' 
\beq
\eta=-\ii\pi(\sigma+\tau)/2\,,
\eeq
is real and positive. 
Define the elliptic gamma-function\footnote{Our function $\Phi(z)$
  coincides with $\Gamma(\EXP^{-2\ii (z-\eta)}; \pp^2,\qq^2)$ in the
  notation of ref. \cite{Spiridonov-essays}. 
The definition \eqref{Phi-def}  
differs slightly from that of ref. \cite{Bazhanov:2010kz}, where the RHS of
\eqref{Phi-def} is denoted as $\Phi(2z)$.}  
\begin{equation}
\Phi(z)\;=\;\prod_{j,k=0}^\infty\frac{1-\EXP^{2\ii
z}\qq^{2j+1}\pp^{2k+1}}{1-\EXP^{-2\ii
z}\qq^{2j+1}\pp^{2k+1}}\;=\;\exp\left\{\sum_{k\neq 0}
\frac{\EXP^{-2\ii
zk}}{k(\qq^k-\qq^{-k})(\pp^k-\pp^{-k})}\right\}\;,\label{Phi-def}
\end{equation}
where the product formula is valid for all $z$, while the 
exponential formula is only valid in the strip
\begin{equation}
-\textrm{Re}\,\eta < \textrm{Im}\,z < \textrm{Re}\,\eta\;.
\end{equation}
The function \eqref{Phi-def} possesses simple periodicity and 
``reflection'' properties
\beq
\Phi(z+\pi)=\Phi(z)\,,
\qquad \Phi(z)\,\Phi(-z)=1\,.\label{reflection}
\eeq
Moreover, it satisfies the following difference equation 
\begin{equation}
\frac{\Phi\left(z-\frac{\pi\sigma}{2}\right)}
{\Phi\left(z+\frac{\pi\sigma}{2}\right)}\;=\;
\prod_{n=0}^\infty (1-\EXP^{2\ii z}\qq^{2n+1})(1-\EXP^{-2\ii
z}\qq^{2n+1})=\overline\vartheta_4(z\,|\,\tau)\;,
\end{equation}
and a similar equation obtained by interchanging 
$\tau$ and $\sigma$. Here
\beq
\overline
\vartheta_j(z\,|\,\tau)=\frac{1}{G(\tau)}\vartheta_j(z\,|\,\tau),\qquad
G(\tau)=\prod_{k=1}^\infty(1-\qq^{2k}),\qquad
\qq=\EXP^{\ii\pi\tau}\,,\label{thetabar} 
\eeq
where $\vartheta_j(z|\tau)$,\  $j=1,2,3,4$, 
 stand for the standard elliptic theta-functions
of the periods $\pi$ and $\pi\tau$, as defined in \cite{WW}.
Note also, than in the regimes \eqref{regimes} the function
\eqref{Phi-def} has a simple complex
conjugation  property,
\beq
\Phi(z)^*=\Phi(-z^*)\,.\label{compconj}
\eeq

\bigskip
\begin{figure}[t]
\centering
\setlength{\unitlength}{1cm}
\begin{picture}(8,6)
\put(1,1){\begin{picture}(7,5)
 \thinlines
 \dottedline{0.08}(1,0)(1,5)\path(0.9,4.9)(1,5)(1.1,4.9)\put(0.9,-0.3){\scriptsize $v'$}
 \dottedline{0.08}(3,0)(3,5)\path(2.9,4.9)(3,5)(3.1,4.9)\put(2.9,-0.3){\scriptsize $v'$}
 \dottedline{0.08}(5,0)(5,5)\path(4.9,4.9)(5,5)(5.1,4.9)\put(4.9,-0.3){\scriptsize $v'$}
 \dottedline{0.08}(0,2)(7,2)\path(6.9,1.9)(7,2)(6.9,2.1)\put(-0.3,1.9){\scriptsize $u'$}
 \dottedline{0.08}(0,4)(7,4)\path(6.9,3.9)(7,4)(6.9,4.1)\put(-0.3,3.9){\scriptsize $u'$}
 \drawline(2,0)(2,5)\path(1.9,4.9)(2,5)(2.1,4.9)\put(1.9,-0.3){\scriptsize $v$}
 \drawline(4,0)(4,5)\path(3.9,4.9)(4,5)(4.1,4.9)\put(3.9,-0.3){\scriptsize $v$}
 \drawline(6,0)(6,5)\path(5.9,4.9)(6,5)(6.1,4.9)\put(5.9,-0.3){\scriptsize $v$}
 \drawline(0,1)(7,1)\path(6.9,0.9)(7,1)(6.9,1.1)\put(-0.3,0.9){\scriptsize $u$}
 \drawline(0,3)(7,3)\path(6.9,2.9)(7,3)(6.9,3.1)\put(-0.3,2.9){\scriptsize $u$}
 \Thicklines
 \path(1.5,0.5)(0.55,1.45)
 \path(0.55,1.55)(1.5,2.5)
 \path(1.5,2.5)(0.55,3.45)
 \path(0.55,3.55)(1.5,4.5)
 \path(1.5,0.5)(2.45,1.45)
 \path(2.45,1.55)(1.5,2.5)
 \path(1.5,2.5)(2.45,3.45)
 \path(2.45,3.55)(1.5,4.5)
 \path(3.5,0.5)(2.55,1.45)
 \path(2.55,1.55)(3.5,2.5)
 \path(3.5,2.5)(2.55,3.45)
 \path(2.55,3.55)(3.5,4.5)
 \path(3.5,0.5)(4.45,1.45)
 \path(4.45,1.55)(3.5,2.5)
 \path(3.5,2.5)(4.45,3.45)
 \path(4.45,3.55)(3.5,4.5)
 \path(5.5,0.5)(4.55,1.45)
 \path(4.55,1.55)(5.5,2.5)
 \path(5.5,2.5)(4.55,3.45)
 \path(4.55,3.55)(5.5,4.5)
 \path(5.5,0.5)(6.45,1.45)
 \path(6.45,1.55)(5.5,2.5)
 \path(5.5,2.5)(6.45,3.45)
 \path(6.45,3.55)(5.5,4.5)
 \put(1.5,0.5){\circle*{0.15}}
 \put(3.5,0.5){\circle*{0.15}}
 \put(5.5,0.5){\circle*{0.15}}
 \put(0.5,1.5){\circle{0.15}}
 \put(2.5,1.5){\circle{0.15}}
 \put(4.5,1.5){\circle{0.15}}
 \put(6.5,1.5){\circle{0.15}}
 \put(1.5,2.5){\circle*{0.15}}
 \put(3.5,2.5){\circle*{0.15}}
 \put(5.5,2.5){\circle*{0.15}}
 \put(0.5,3.5){\circle{0.15}}
 \put(2.5,3.5){\circle{0.15}}
 \put(4.5,3.5){\circle{0.15}}
 \put(6.5,3.5){\circle{0.15}}
 \put(1.5,4.5){\circle*{0.15}}
 \put(3.5,4.5){\circle*{0.15}}
 \put(5.5,4.5){\circle*{0.15}}
 \end{picture}}
 \end{picture}
 \caption{The square lattice shown with bold sites and bold edges drawn
   diagonally. The associated medial lattice is drawn with thin and
   dotted horizontal and vertical lines. The lines are oriented and
   carry rapidity variables $u$, $u'$, $v$ and $v'$.}
\label{fig-lattice}
\end{figure}
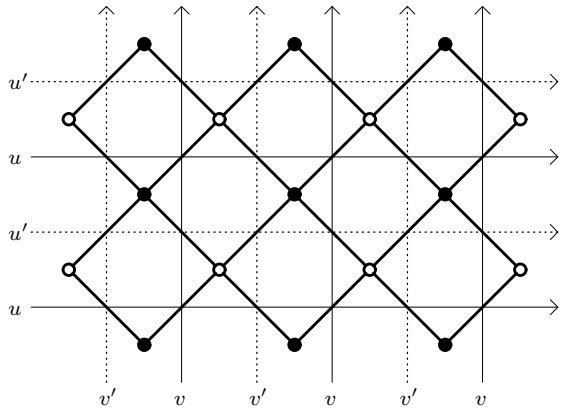
Next, we want to introduce a new two-dimensional solvable edge-interaction  
model.   
The model can be formulated on rather general planar
graphs, however, for the purposes of this presentation it is convenient to
take a regular square lattice. 
Consider the square lattice, drawn diagonally as in
Fig.~\ref{fig-lattice}.      
The edges of the lattice are shown with bold lines 
and the sites are shown with either open or filled circles in a
checkerboard order. 
In this Section we will not distinguish the two type sites; 
their difference will be important in Sect.~\ref{sec:ybe}. At each lattice  site place a $n$-component continuous spin variable 
\begin{equation}
\boldsymbol{x}=\{x_1,\dots,x_n\}\in\mathbb{R}^n, \qquad 
0\le x_j< \pi,\qquad \sum_{j=1}^n x_j = 0\pmod \pi\;. \label{spinvar}
\end{equation}
Note that due to the restriction on the total sum, there are only
$(n-1)$ independent variables $x_j$. For further reference define the
integration measure
\begin{equation}\label{measure}
\int d\boldsymbol{x}=\int_0^{\pi}\cdots \int_0^{\pi} dx_1\cdots
dx_{n-1}\;,\quad 
\mathds{1}\;=\;\int |\boldsymbol{x}\rangle d\boldsymbol{x}
\langle\boldsymbol{x}|\;. 
\end{equation}
Fig.~\ref{fig-lattice} also shows an auxiliary {\em medial}\/ graph 
whose sites lie on the edges of the original square lattice. 
The medial graph is drawn with alternating 
thin and dotted lines. The lines are directed as indicated by  arrows.     
To each horizontal (vertical) line on the medial graph assign 
a rapidity variable $u$ ($v$). In general these variables may be 
different for different lines. However, a convenient level
of generality that we shall use here is 
to assign the same rapidity $u$ to all thin horizontal lines 
and the same variable $u'$ to all dotted horizontal lines. Similarly,
assign the variables $v$ and $v'$ to thin and dotted 
vertical lines as indicated in 
Fig.~\ref{fig-lattice}.
\begin{figure}[ht]
\begin{center}
\setlength{\unitlength}{1cm}
\begin{picture}(15,4)
\put(0.5,1.5)
{\begin{picture}(6,2)
 \thinlines
 \drawline(0,0)(2,2)\path(2,1.9)(2,2)(1.9,2)
 \put(0,-0.3){\scriptsize $u$}
 \dottedline{0.08}(2,0)(0,2)\path(0,1.9)(0,2)(0.1,2)
 \put(2,-0.3){\scriptsize $v'$}
 \Thicklines
 \path(0,1)(2,1)
 \put(0,1){\circle*{0.1}}
 \put(2,1){\circle*{0.1}}
 \put(-0.4,0.9){\scriptsize $x$}
 \put(2.2,0.9){\scriptsize $y$}
 \put(0.3,-1.0){\scriptsize $\iW_{u-v'}(\boldsymbol{x},\boldsymbol{y})$}
 \thinlines
 \drawline(4,0)(6,2)\path(6,1.9)(6,2)(5.9,2)
 \drawline(6,0)(4,2)\path(4,1.9)(4,2)(4.1,2)
 \put(4,-0.3){\scriptsize $u$}
 \put(6,-0.3){\scriptsize $v$}
 \Thicklines
 \path(5,0)(5,2)
 \put(5,0){\circle*{0.1}}
 \put(5,2){\circle*{0.1}}
 \put(4.9,-0.3){\scriptsize $y$}
 \put(4.9,2.2){\scriptsize $x$}
 \put(4.3,-1.0){\scriptsize $\overline{\iW}_{u-v}(\boldsymbol{y},\boldsymbol{x})$}
\end{picture}}
\put(8,1.5){\begin{picture}(6,2)
 \thinlines
 \dottedline{0.08}(0,0)(2,2)\path(2,1.9)(2,2)(1.9,2)
 \put(0,-0.3){\scriptsize $u'$}
 \drawline(2,0)(0,2)\path(0,1.9)(0,2)(0.1,2)
 \put(2,-0.3){\scriptsize $v$}
 \Thicklines
 \path(0,1)(2,1)
 \put(0,1){\circle*{0.1}}
 \put(2,1){\circle*{0.1}}
 \put(-0.4,0.9){\scriptsize $y$}
 \put(2.2,0.9){\scriptsize $x$}
 \put(0.3,-1.0){\scriptsize $\iW_{u'-v}(\boldsymbol{x},\boldsymbol{y})$}
 \thinlines
 \dottedline{0.08}(4,0)(6,2)\path(6,1.9)(6,2)(5.9,2)
 \dottedline{0.08}(6,0)(4,2)\path(4,1.9)(4,2)(4.1,2)
 \put(4,-0.3){\scriptsize $u'$}
 \put(6,-0.3){\scriptsize $v'$}
 \Thicklines
 \path(5,0)(5,2)
 \put(5,0){\circle*{0.1}}
 \put(5,2){\circle*{0.1}}
 \put(4.9,-0.3){\scriptsize $x$}
 \put(4.9,2.2){\scriptsize $y$}
 \put(4.3,-1.0){\scriptsize $\overline{\iW}_{u'-v'}(\boldsymbol{y},\boldsymbol{x})$}
\end{picture}}
\end{picture}
\caption{Four different types of edges and their Boltzmann weights.}
\label{fig-crosses}
\end{center}
\end{figure}
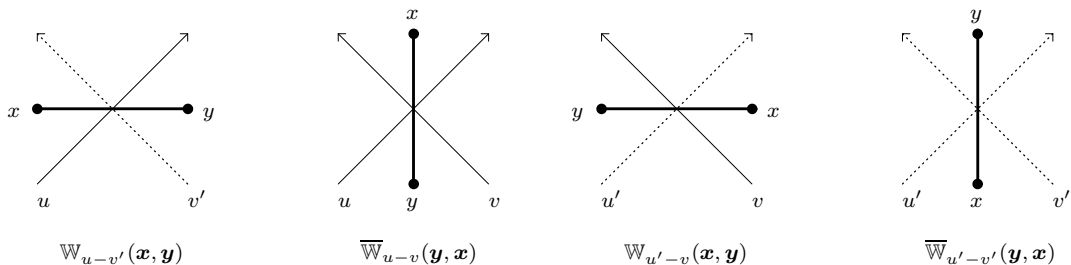

Two spins interact only if they are connected with an edge.
The corresponding Boltzmann weight depends
on spins at the ends of the edge and on two rapidities passing
through the edge.  There are four types of edges differing by
orientations and types of the directed rapidity lines passing through
the edge. They are assigned with different Boltzmann weights
as shown in Fig.\ref{fig-crosses}. The weights are defined as
\begin{equation}
\iW_\alpha(\boldsymbol{x},\boldsymbol{y})
=\kappa_n(\alpha)^{-1}\prod_{j,k=1}^{n}\Phi(x_j-y_k+\ii\alpha),\quad  
\overline{\iW}_\alpha(\boldsymbol{x},\boldsymbol{y})
=\sqrt{\iS(\boldsymbol{x})\iS(\boldsymbol{y})}\iW_{\eta-\alpha}(x,y)\;,
\label{weights}
\end{equation}
where the single-spin function $\iS$ is given by
\begin{equation}
\iS(\boldsymbol{x})\;=\;\varkappa_s^{-1}
\ds\prod_{j\neq k}
  \Big\{\Phi(x_j-x_k+\ii\eta)\Big\}^{-1}=
\varkappa_s^{-1}\ds\prod_{j<k} \Big\{\EXP^{\eta/2}\,
\overline{\vartheta}_1(x_j-x_k\,|\,\tau)\,
\overline{\vartheta}_1(x_j-x_k\,|\,\sigma)\Big\}\,,\label{S-def}
\eeq
and
\beq
\varkappa_s\;=\;{n!}\left(\frac{\pi}
{G(\tau)\,G(\sigma)}\right)^{n-1}\;. \label{ks-def}
\end{equation}
Here the indices $j,k$ run over the values $1,2,\ldots,n$; the functions
$\overline\vartheta(x\,|\,\tau)$ and $G(\tau)$ are defined in \eqref{thetabar}.
The normalization factor 
\begin{equation}
\kappa_n(\alpha)\;=\;\exp\left\{\sum_{k\neq 0}
\frac{\EXP^{2nk\alpha}}{k\;(\pp^k-\pp^{-k})(\qq^k-\qq^{-k})}\;
\frac{\pp^k\qq^k-\pp^{-k}\qq^{-k}}
{\pp^{nk}\qq^{nk}-\pp^{-nk}\qq^{-nk}}\right\}\;,\label{pf-edge}
\end{equation}
has the meaning of the partition function per edge for unnormalized
Boltzmann weights (i.e., when the factor $\kappa_n(\alpha)$ in
\eqref{weights} is omitted). 
It solves a pair of functional equations
\begin{equation}\label{feq}
\kappa_n(\alpha)\, \kappa_n(-\alpha)=1,\quad
\kappa_n(\eta-\alpha)\,\kappa_n(\eta+\alpha)=
\Phi(\ii\eta-\ii n \alpha)\,\Phi(\ii\eta+\ii n\alpha)\;.
\end{equation}
The weights \eqref{weights} satisfy two inversion relations:
\begin{equation}\label{inversion}
\iW_\alpha(\boldsymbol{x},\boldsymbol{y})\,
\iW_{-\alpha}(\boldsymbol{y},\boldsymbol{x})=1\;,\quad
\int d\boldsymbol{x} \,\overline{\iW}_\alpha(\boldsymbol{x},\boldsymbol{z})\,
\overline{\iW}_{-\alpha}(\boldsymbol{z},\boldsymbol{y})
=\frac{1}{n!}\sum_{\hat\sigma}\delta(\boldsymbol{x},\hat\sigma(\boldsymbol{y}))
\end{equation}
where the sum is taken over $n!$ permutations $\hat\sigma$ of components of the
vector $\boldsymbol{y}=\{y_1,\dots,y_n\}$ and the 
$\delta$-function is understood
with respect to the measure \eqref{measure}. The first of these
relations is a trivial corollary of the definition \eqref{weights} and
the reflection property of the elliptic gamma function $\Phi(z)$. The
second relation is a particular case of a more general relation stated
in Theorem~11 in ref.\cite{Spiridonov-essays}.

Note that for any permutation $\hat\sigma$ one has 
\begin{equation}
\iW(\boldsymbol{x},\boldsymbol{y})=
\iW_\alpha(\hat\sigma(\boldsymbol{x}),\boldsymbol{y})=\iW(\boldsymbol{x},\hat\sigma(\boldsymbol{y}))\;,
\quad \iS(\boldsymbol{x})=\iS(\hat\sigma(\boldsymbol{x}))\;,\label{ws-sym}
\end{equation}
as a trivial consequence of the definitions \eqref{weights} and \eqref{S-def}.
Under the complex conjugation the weights transform as 
\begin{equation}
\iW_\alpha(\boldsymbol{x},\boldsymbol{y})^*=\iW_{\alpha^*}
(\boldsymbol{y},\boldsymbol{x})
=\iW_{\alpha^*}(-\boldsymbol{x},-\boldsymbol{y})\;,\quad
\iS(\boldsymbol{x})^*=\iS(\boldsymbol{x})\;.\label{ws-conj}
\end{equation}
provided the nomes $\pp$ and $\qq$ 
belong to either of the regimes, defined in \eqref{regimes} 
(remind that the spin
variables are always assumed to be real; see \eqref{spinvar}). 
When $n=2$ the weights $\iW$ and $\iS$ are real and positive
\cite{Bazhanov:2010kz}. Correspondingly, 
the weights $\iW_\alpha(\boldsymbol{x},\boldsymbol{y})$ 
and $\overline{\iW}_\alpha(\boldsymbol{x},\boldsymbol{y})$ 
are symmetric under an exchange of the spins 
$\boldsymbol{x}$ and $\boldsymbol{y}$. 
However, for $n>2$ this symmetry is lost. As a particular consequence
of this fact the above description of 
the associated lattice model required two types of the rapidity lines. 
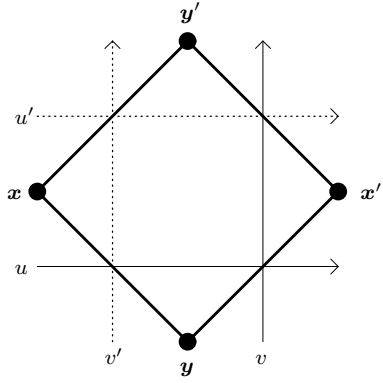
\begin{figure}[ht]
\begin{center}
\setlength{\unitlength}{1cm}
\begin{picture}(5,5.4)
\put(1,.5){\begin{picture}(4,4)
 \thinlines
 \drawline(3,0)(3,4)\path(2.9,3.9)(3,4)(3.1,3.9)
 \put(2.9,-0.3){\scriptsize $v$}
 \drawline(0,1)(4,1)\path(3.9,0.9)(4,1)(3.9,1.1)
 \put(-0.3,0.9){\scriptsize $u$}
 \dottedline{0.08}(1,0)(1,4)\path(0.9,3.9)(1,4)(1.1,3.9)
 \put(0.9,-0.3){\scriptsize $v'$}
 \dottedline{0.08}(0,3)(4,3)\path(3.9,2.9)(4,3)(3.9,3.1)
 \put(-0.3,2.9){\scriptsize $u'$}
 \Thicklines
 \path(2,0)(4,2)
 \path(4,2)(2,4)
 \path(2,4)(0,2)
 \path(0,2)(2,0)
 \put(2,0){\circle*{0.2}}
 \put(4,2){\circle*{0.2}}
 \put(2,4){\circle*{0.2}}
 \put(0,2){\circle*{0.2}}
 \put(1.9,-0.4){\scriptsize $\boldsymbol{y}$}
 \put(1.9,4.3){\scriptsize $\boldsymbol{y}'$}
 \put(-0.4,1.9){\scriptsize $\boldsymbol{x}$}
 \put(4.3,1.9){\scriptsize $\boldsymbol{x}'$}
 \end{picture}}
\end{picture} 
\caption{Pictorial representation of the $R$-matrix
 $\langle\boldsymbol{x},\boldsymbol{y}|R_{\boldsymbol{uv}}|\boldsymbol{x}',\boldsymbol{y}'\rangle$.}\label{fig-box}
\end{center}
\end{figure}

The square lattice in Fig. \ref{fig-lattice} can be  formed by periodic
translations of the ``box diagram'' shown in Fig.~\ref{fig-box}.
The Boltzmann weight of this box 
can be conveniently associated with an $R$-matrix,
\begin{equation}\label{bigR}
\langle 
\boldsymbol{x},\boldsymbol{y}|\mathbb{R}_{\boldsymbol{u}\boldsymbol{v}}|
\boldsymbol{x}',\boldsymbol{y}'\rangle =
\overline{\iW}_{u-v}(\boldsymbol{y},\boldsymbol{x}')
\overline{\iW}_{u'-v'}(\boldsymbol{y}',\boldsymbol{x})
\iW_{u'-v}(\boldsymbol{x}',\boldsymbol{y}')
\iW_{u-v'}(\boldsymbol{x},\boldsymbol{y})\;,
\end{equation} 
where $\boldsymbol{u}=[u,u']$ and 
$\boldsymbol{q}=[v,v']$ 
stand for the pairs of rapidities in horizontal and vertical
directions.
The partition function is
defined as 
\beq
Z=\int  \ \prod_{\rm{boxes}}
\ 
\langle \boldsymbol{x},\boldsymbol{y}\,|\,\mathbb{R}_{\boldsymbol{u}
\boldsymbol{v}}\,|\,\boldsymbol{x}',\boldsymbol{y}'\rangle
\ \prod_{\rm sites} d \boldsymbol{x}, \label{Z-vertex}
\eeq
where the first product is taken over all boxes and $\boldsymbol{x},
\boldsymbol{y},\boldsymbol{x}',\boldsymbol{y}'$ are the spins at the
corners arranged as in Fig.~\ref{fig-box}. The integral is
taken over all configurations of the spin variables on the internal
lattice sites. The boundary spins are kept fixed. Note that due to 
periodicity \eqref{reflection} the definition \eqref{Z-vertex}
only contains closed contour integrals (the contours can
be deformed into the complex plane, if necessary).

\section{Yang-Baxter equations and star-star relations}\label{sec:ybe}
The $R$-matrix \eqref{bigR} can be regarded as a kernel of an integral
operator acting on a pair of (multi-component) continuous spin
variables \eqref{spinvar}. We claim that it satisfies the Yang-Baxter
equation of the form 
\begin{equation}
\begin{array}{l}
\ds \int d\boldsymbol{x}'d\boldsymbol{y}'d\boldsymbol{z}'
\langle\boldsymbol{x},\boldsymbol{y}|\iR_{\boldsymbol{u}\boldsymbol{v}}|\boldsymbol{x}',\boldsymbol{y}'\rangle
\langle\boldsymbol{x}',\boldsymbol{z}|\iR_{\boldsymbol{u}\boldsymbol{w}}|\boldsymbol{x}'',\boldsymbol{z}'\rangle
\langle\boldsymbol{y}',\boldsymbol{z}'|\iR_{\boldsymbol{v}\boldsymbol{w}}|\boldsymbol{y}'',\boldsymbol{z}''\rangle
=\\
[5mm]
\phantom{xxxxxxxxx}
\ds \int d\boldsymbol{x}'d\boldsymbol{y}'d\boldsymbol{z}'
\langle\boldsymbol{y},\boldsymbol{z}|\iR_{\boldsymbol{v}\boldsymbol{w}}|\boldsymbol{y}',\boldsymbol{z}'\rangle
\langle\boldsymbol{x},\boldsymbol{z}'|\iR_{\boldsymbol{u}\boldsymbol{w}}|\boldsymbol{x}',\boldsymbol{z}''\rangle
\langle\boldsymbol{x}',\boldsymbol{y}'|\iR_{\boldsymbol{u}\boldsymbol{v}}|\boldsymbol{x}'',\boldsymbol{y}''\rangle\;,
\end{array}\label{YBE-int}
\end{equation}
where the integration measure is defined in \eqref{measure}, and the
symbols 
$\boldsymbol{u}=[u,u']$, $\boldsymbol{v}=[v,v']$ and
$\boldsymbol{w}=[w,w']$
stand for the rapidity pairs. With the standard conventions the last
equation can be written in an operator form 
\begin{equation}
\iR_{\boldsymbol{u}\boldsymbol{v}}\,
\iR_{\boldsymbol{u}\boldsymbol{w}}\,
\iR_{\boldsymbol{v}\boldsymbol{w}}
=
\iR_{\boldsymbol{v}\boldsymbol{w}}\,
\iR_{\boldsymbol{u}\boldsymbol{w}}\,
\iR_{\boldsymbol{u}\boldsymbol{v}}\;. \label{YBE-op}
\end{equation}
The $R$-matrix \eqref{bigR} was derived as a unique intertwiner 
for two different sets of Lax operators serving the modular double
\cite{Faddeev:1999,Spiridonov-essays}  
of the $sl(n)$ analog of the quadratic Sklyanin algebra
\cite{Belavin:1981ix,Skl82}. The details of calculations will be
published elsewhere. 

\begin{figure}[ht]
\begin{center}
\setlength{\unitlength}{1cm}
\begin{picture}(10,4)
\put(1,1){\begin{picture}(3,3)
 \thinlines
 \drawline(0.5,-0.5)(0.5,2.5)\path(0.4,2.4)(0.5,2.5)(0.6,2.4)
 \drawline(-0.5,0.5)(2.5,0.5)\path(2.4,0.4)(2.5,0.5)(2.4,0.6)
 \put(0.4,-0.8){\scriptsize $v$}
 \put(-0.8,0.4){\scriptsize $u$}
 \dottedline{0.08}(1.5,-0.5)(1.5,2.5)\path(1.4,2.4)(1.5,2.5)(1.6,2.4)
 \dottedline{0.08}(-0.5,1.5)(2.5,1.5)\path(2.4,1.4)(2.5,1.5)(2.4,1.6)
 \put(1.4,-0.8){\scriptsize $v'$}
 \put(-0.8,1.4){\scriptsize $u'$}
 \Thicklines
\path(0,0)(.95,.95)
 \path(1.05,1.05)(2,2)
 \path(1.05,.95)(2,0)
 \path(.95,1.05)(0,2)
 \put(1,1){\circle {0.15}}
 \put(0,0){\circle*{0.15}}
 \put(0,2){\circle*{0.15}}
 \put(2,0){\circle*{0.15}}
 \put(2,2){\circle*{0.15}}
 \put(0.9,1.2){\scriptsize $\boldsymbol{x}$}
 \put(-0.2,2.2){\scriptsize $\boldsymbol{a}$}
 \put(-0.2,-0.4){\scriptsize $\boldsymbol{c}$}
 \put(2.1,2.2){\scriptsize $\boldsymbol{b}$}
 \put(2.1,-0.4){\scriptsize $\boldsymbol{d}$}
\end{picture}}
\put(6,1){\begin{picture}(3,3)
 \thinlines
 \dottedline{0.08}(0.5,-0.5)(0.5,2.5)\path(0.4,2.4)(0.5,2.5)(0.6,2.4)
 \dottedline{0.08}(-0.5,0.5)(2.5,0.5)\path(2.4,0.4)(2.5,0.5)(2.4,0.6)
 \put(0.4,-0.8){\scriptsize $v'$}
 \put(-0.8,0.4){\scriptsize $u'$}
 \drawline(1.5,-0.5)(1.5,2.5)\path(1.4,2.4)(1.5,2.5)(1.6,2.4)
 \drawline(-0.5,1.5)(2.5,1.5)\path(2.4,1.4)(2.5,1.5)(2.4,1.6)
 \put(1.4,-0.8){\scriptsize $v$}
 \put(-0.8,1.4){\scriptsize $u$}
 \Thicklines
 \path(0.05,0.05)(.95,.95)
 \path(1.05,1.05)(1.95,1.95)
 \path(1.05,.95)(1.95,0.05)
 \path(.95,1.05)(0.05,1.95)
 \put(1,1){\circle*{0.15}}
 \put(0,0){\circle{0.15}}
 \put(0,2){\circle{0.15}}
 \put(2,0){\circle{0.15}}
 \put(2,2){\circle{0.15}}
 \put(0.9,1.2){\scriptsize $\boldsymbol{y}$}
 \put(-0.2,2.2){\scriptsize $\boldsymbol{a}$}
 \put(-0.2,-0.4){\scriptsize $\boldsymbol{c}$}
 \put(2.1,2.2){\scriptsize $\boldsymbol{b}$}
 \put(2.1,-0.4){\scriptsize $\boldsymbol{d}$}
\end{picture}}
\end{picture}
\caption{Two types of four-edge stars: a white-centred star
  $\mathbb{V}^{(1)}$ (left) and a black-centred 
  $\mathbb{V}^{(2)}$ (right)}\label{fig-IRF} 
\end{center}
\end{figure}
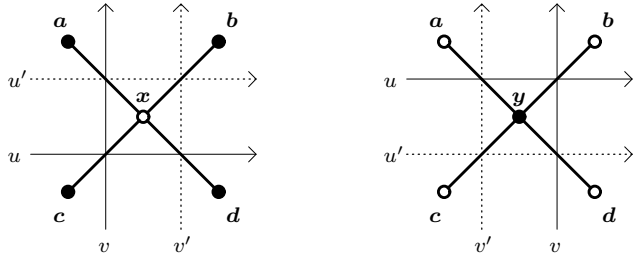

According to the terminology of the Baxter's book 
\cite{Baxterbook}, Eq.\eqref{Z-vertex}
defines a {\em vertex\/} model. It is easy to see that to within
boundary effects the same model can be equivalently reformulated
as an {\em interaction-round-a-face\/} (IRF) model. Indeed, the
lattice in Fig.~\ref{fig-lattice} can also be formed by periodic
translations of a {\em four-edge star} (consisting of 
four edges meeting at the same site), 
instead of the box diagram of Fig.~\ref{fig-box}. 
A little  inspection shows that 
there are only two different types of
such stars shown in Fig.~\ref{fig-IRF}. 
Recall that there are types of sites, shown with open and filled
circles. We will refer to them as to ``white'' or ``black'' sites,
respectively. 
There are white-centred stars (i.e., centred around a white sites) and
black-centred ones.
Applying the rules shown in Fig.~\ref{fig-crosses} one can write 
IRF-type Boltzmann weights corresponding to these stars 
\begin{equation}\label{V1}
\ds \mathbb{V}_{\boldsymbol{u}\boldsymbol{v}}^{(1)}\left(\begin{array}{cc}
\boldsymbol{a} & \boldsymbol{b}\\ \boldsymbol{c} & \boldsymbol{d}\end{array}\right)=
\int d\boldsymbol{x}\,
\overline{\iW}_{u-v}(\boldsymbol{c},\boldsymbol{x})\,
\overline{\iW}_{u'-v'}(\boldsymbol{b},\boldsymbol{x})\,
\iW_{u'-v}(\boldsymbol{x},\boldsymbol{a})\,
\iW_{u-v'}(\boldsymbol{x},\boldsymbol{d})\;,
\end{equation}
and
\begin{equation}\label{V2}
\ds \mathbb{V}_{\boldsymbol{u}\boldsymbol{v}}^{(2)}\left(\begin{array}{cc}
\boldsymbol{a} & \boldsymbol{b}\\ \boldsymbol{c} &
\boldsymbol{d}\end{array}\right)=\int d\boldsymbol{y}\, 
\overline{\iW}_{u-v}(\boldsymbol{y},\boldsymbol{b})\,  
\overline{\iW}_{u'-v'}(\boldsymbol{y},\boldsymbol{c})\,
\iW_{u'-v}(\boldsymbol{d},\boldsymbol{y})\,
\iW_{u-v'}(\boldsymbol{a},\boldsymbol{y})\;,
\end{equation}
where the bold symbols $\boldsymbol{u}$ and $\boldsymbol{v}$ has the
same meaning as in \eqref{YBE-int}.
The above two expressions are connected by the so-called {\em star-star\/}
relation \cite{Baxter:1986phd,Bazhanov:1992jqa}, which in our case reads 
\begin{equation}
{\iW_{v'-v}(\boldsymbol{d},\boldsymbol{c})\,
\iW_{u'-u}(\boldsymbol{d},\boldsymbol{b})} 
\,\,\mathbb{V}_{\boldsymbol{u}\boldsymbol{v}}^{(1)}\left(\begin{array}{cc}
\boldsymbol{a} & \boldsymbol{b} \\
\boldsymbol{c} & \boldsymbol{d}\end{array}\right)\;=\;
{\iW_{v'-v}(\boldsymbol{b},\boldsymbol{a})\,
\iW_{u'-u}(\boldsymbol{c},\boldsymbol{a})}
\,\,
\mathbb{V}_{\boldsymbol{u}\boldsymbol{v}}^{(2)}\left(\begin{array}{cc}
\boldsymbol{a} & \boldsymbol{b} \\
\boldsymbol{c} & \boldsymbol{d}\end{array}\right)\;.\label{star-star}
\end{equation}
Apparently this is the simplest relation for the Boltzmann weights 
which ensures the integrability of the considered model\footnote{%
\label{foot3}%
For $n = 2$ the star-star relation \eqref{star-star} is just a
consequence of the star-triangle relation, Eq.(1.5) of
\cite{Bazhanov:2010kz}, which is equivalent to the elliptic beta
integral \cite{Spiridonov:2010em,Spiridonov-beta}.  However, for $n\ge3$ the
corresponding star-triangle relation apparently does not exist (at
least it is not known to the authors) and the star-star relation
\eqref{star-star} seems to be the simplest relation of this type.}.
Currently, we claim this relation as a conjecture,
however, we expect that a complete proof could be obtained by a rather
straightforward generalization of the results of
\cite{Bazhanov:1990qk,Date:1990bs,Kashaev:1990wy,Bazhanov:1992jqa,Znbrok}, 
devoted to
discrete-spin models connected with the $sl(n)$ algebra. 
We have verified this relation in a few orders of
perturbation theory 
in the temperature-like variables (see Appendix~\ref{appB}) and made
extensive numerical checks for $n=3,4$.

It is worth noting that the
Yang-Baxter equation \eqref{YBE-int} for the composite ``box
$R$-matrix'' \eqref{bigR} is a simple corollary of the star-star
relation \eqref{star-star}.
The same relation also implies yet another Yang-Baxter equation for the
IRF-type weights  
\begin{equation}
\begin{array}{l}
\ds \int d\boldsymbol{h}\;
\mathbb{V}_{\boldsymbol{uv}}\left(\begin{array}{cc}
\boldsymbol{c} & \boldsymbol{h} \\
\boldsymbol{e} & \boldsymbol{d}\end{array}\right)
\mathbb{V}_{\boldsymbol{uw}}\left(\begin{array}{cc}
\boldsymbol{h} & \boldsymbol{b} \\
\boldsymbol{d} & \boldsymbol{f}\end{array}\right)
\mathbb{V}_{\boldsymbol{vw}}\left(\begin{array}{cc}
\boldsymbol{c} & \boldsymbol{g} \\
\boldsymbol{h} & \boldsymbol{b}\end{array}\right)=\\
[5mm]
\ds \phantom{xxxxxxxxxxxxxxxxx}
\int d\boldsymbol{a}\;
\mathbb{V}_{\boldsymbol{vw}}\left(\begin{array}{cc}
\boldsymbol{e} & \boldsymbol{a} \\
\boldsymbol{d} & \boldsymbol{f}\end{array}\right)
\mathbb{V}_{\boldsymbol{uw}}\left(\begin{array}{cc}
\boldsymbol{c} & \boldsymbol{g} \\
\boldsymbol{e} & \boldsymbol{a}\end{array}\right)
\mathbb{V}_{\boldsymbol{uv}}\left(\begin{array}{cc}
\boldsymbol{g} & \boldsymbol{b} \\
\boldsymbol{a} & \boldsymbol{f}\end{array}\right)
\end{array}\label{YBE-IRF}
\end{equation}
where 
\begin{equation}
\begin{array}{rcl}
\mathbb{V}_{\boldsymbol{u}\boldsymbol{v}}\left(\begin{array}{cc}
\boldsymbol{a} & \boldsymbol{b} \\
\boldsymbol{c} & \boldsymbol{d}\end{array}\right)&=&
\left\{\frac
{\ds
\iW_{v'-v}(\boldsymbol{d},\boldsymbol{c})
\iW_{u'-u}(\boldsymbol{d},\boldsymbol{b})}
{\ds\iW_{v'-v}(\boldsymbol{b},\boldsymbol{a})  
\iW_{u'-u}(\boldsymbol{c},\boldsymbol{a})}
\right\}^{\frac{1}{2}}
\ 
\mathbb{V}_{\boldsymbol{u}\boldsymbol{v}}^{(1)}\left(\begin{array}{cc}
\boldsymbol{a} & \boldsymbol{b} \\
\boldsymbol{c} & \boldsymbol{d}\end{array}\right)
\\[7mm]
&=&
\left\{{\frac{\ds\iW_{v'-v}(\boldsymbol{b},\boldsymbol{a})  
\iW_{u'-u}(\boldsymbol{c},\boldsymbol{a})}
{\ds
\iW_{v'-v}(\boldsymbol{d},\boldsymbol{c}) 
\iW_{u'-u}(\boldsymbol{d},\boldsymbol{b})}}
\right\}^{\frac{1}{2}}\ 
\mathbb{V}_{\boldsymbol{u}\boldsymbol{v}}^{(2)}\left(\begin{array}{cc}
\boldsymbol{a} & \boldsymbol{b} \\
\boldsymbol{c} & \boldsymbol{d}\end{array}\right)\,.
\end{array}
\label{weight-IRF}
\end{equation}

The partition function \eqref{Z-vertex} can be re-written using either
of the weights \eqref{V1}, \eqref{V2} or \eqref{weight-IRF}. For
example, it easy to see that, up to boundary effects, 
\beq
Z=\int  \ \prod_{\rm (white\  stars)}
\ 
\mathbb{V}_{\boldsymbol{u}\boldsymbol{v}}\left(\begin{array}{cc}
\boldsymbol{a} & \boldsymbol{b} \\
\boldsymbol{c} & \boldsymbol{d}\end{array}\right)
\ \prod_{\rm (black\  sites)} d \boldsymbol{x}, \label{Z-IRF}
\eeq
where the first product is taken over all white-centred stars and
$\boldsymbol{a},
\boldsymbol{b},\boldsymbol{c},\boldsymbol{d}$ denote the corresponding
outer spins, arranged as in Fig.~\ref{fig-IRF}. 
The integral is
taken over the spin variables on black internal
sites (the integration over the spins on white sites is included through the
definitions \eqref{weight-IRF} and  \eqref{V1}). As before, 
the boundary spins are kept fixed. Note that 
the weights \eqref{V1}, \eqref{V2} and \eqref{weight-IRF} differ from
each other by 
equivalence transformations which leave the partition function \eqref{Z-IRF}
unchanged (up to boundary contributions). 

There are two alternative forms for the RHS in 
\eqref{weight-IRF}, which coincide by virtue of the star-star relation 
\eqref{star-star}. This fact 
allows one to understand the behaviour of the
IRF-weight \eqref{weight-IRF} under the complex conjugation.  
Consider any of the regimes defined in \eqref{regimes}
and assume that the rapidity
variables obey the relations 
\begin{equation}
u^*=u'\;,\;\;
v^*=v'\;,\quad\textrm{and}\quad
0<\textrm{Re}(u'-v',
u-v, \dots)<\eta\;.\label{pq-rel}
\end{equation}
Note that 
these relations allow a homogeneous case $u=u'$, \ $v=v'$, when $u$ and $v$
are real. Taking \eqref{pq-rel} into account and using
\eqref{ws-conj} one can easily check that the complex conjugation just
interchanges the two equivalent forms of \eqref{weight-IRF}.
It follows then that the 
IRF weight \eqref{weight-IRF} is real.
Further, one can show that there must exist a region for the
parameters $\pp$ and $\qq$ where this weight is non-negative, 
\begin{equation}
\left[\mathbb{V}_{\boldsymbol{u}\boldsymbol{v}}\left(\begin{array}{cc}
\boldsymbol{a} & \boldsymbol{b} \\
\boldsymbol{c} & \boldsymbol{d}\end{array}\right)\right]^*\;=\;
\mathbb{V}_{\boldsymbol{u}\boldsymbol{v}}\left(\begin{array}{cc}
\boldsymbol{a} & \boldsymbol{b} \\
\boldsymbol{c} & \boldsymbol{d}\end{array}\right)\ge0\;,\label{positive}
\end{equation}
for all values of the spin variables. 
First of all 
note, that in the regimes \eqref{regimes}
the single-spin weight \eqref{S-def} is real and
non-negative
\beq
\iS(\boldsymbol{x})\ge0\,.
\eeq
It vanishes only then at least two components of $\boldsymbol{x}$
coincide.
Next, the definition \eqref{weights} for
$\overline{\iW}_{\alpha}(\boldsymbol{x},\boldsymbol{y})$  contains
square roots of $\iS(\boldsymbol{x})$ and $\iS(\boldsymbol{y})$ and
therefore there is an ambiguity in choosing signs. 
However, it is easy to see that these signs cancel out for all
internal sites and therefore can be chosen arbitrarily. For definiteness
we assume that $\iS(\boldsymbol{x})^\frac{1}{2}=
|\iS(\boldsymbol{x})^\frac{1}{2}|\ge0$.

Consider the limit when $\pp,\qq\to0$ and the ratio $\pp/\qq$ is
finite. It is convenient to parametrize the rapidity variables as  
\beq
\begin{array}{rclrcl}
u&=&\ds\frac{\eta}{2}+\alpha+\frac{\ii}{2}(\gamma-\beta),\qquad&
v&=&\ds-\frac{\ii}{2}(\gamma+\beta),\\[5mm]
u'&=&\ds\frac{\eta}{2}+\alpha-\frac{\ii}{2}(\gamma-\beta),&
v'&=&\ds+\frac{\ii}{2}(\gamma+\beta)
\end{array}\label{param} 
\eeq
where the variables $\alpha$, $\beta$ and $\gamma$ are real. 
Moreover, assume that 
\beq
|{\rm Re}\alpha|\ll \eta\,,
\eeq
which means that $\EXP^{2\alpha}={\mathcal O}(1)$, when 
$\pp,\qq\to0$.
The relations \eqref{pq-rel} are automatically satisfied. The
weights \eqref{weights} and \eqref{S-def} can be expanded in powers of
$\pp$ and $\qq$ 
\beq
\begin{array}{rcl}
\iW_\alpha(\boldsymbol{x},\boldsymbol{y})&=&1+{\mathcal
  O}(\pp\qq)+\ldots\,,\\[6mm]
\iW_{\eta/2+\alpha}(\boldsymbol{x},\boldsymbol{y})&=&1+{\mathcal
  O}((\pp\qq)^\hf)+{\mathcal
  O}(\pp\qq)+\ldots\,,
\\[.6cm]\ds
{\pi^{n-1}n!}\ \iS(\boldsymbol{x})\,\prod_{j<k}
  \big(2\sin(x_j-x_k)\big)^{-2}&=&\ds
1+{\mathcal O}(\pp^2+\qq^2)+\ldots\,.
\end{array}\label{WSexp}
\eeq
Explicit form of the coefficients to within the forth order is given
in Appendix~\ref{appB} (the fractional powers arise due to the relation
$\EXP^{-\eta}=(\pp\qq)^\hf$). 
It is convenient to define sums of exponents of the spin
variables, entering the IRF weight \eqref{weight-IRF}  
\begin{equation}
{\mathcal A}_1=\sum_{j=1}^n\EXP^{2\ii a_j},\quad
{\mathcal B}_1=\sum_{j=1}^n\EXP^{2\ii b_j},\quad
{\mathcal C}_1=\sum_{j=1}^n\EXP^{2\ii c_j},\quad 
{\mathcal D}_1=\sum_{j=1}^n\EXP^{2\ii d_j},
\end{equation}
Using the above expansions one obtains
\beq
{\Big|\,\iS(\boldsymbol{b})^{-\frac{1}{2}}
\,\iS(\boldsymbol{c})^{-\frac{1}{2}}\,\Big|^{\phantom{|}}}\ 
\mathbb{V}_{\boldsymbol{u}\boldsymbol{v}}\left(\begin{array}{cc}
\boldsymbol{a} & \boldsymbol{b} \\
\boldsymbol{c} & \boldsymbol{d}\end{array}\right)=
1+\pp\qq\  {\mathcal P}+{\mathcal
  O}((\pp\qq)^{\frac{3}{2}})+\ldots\label{Vexp}
\eeq
where ${\mathcal P}\ge0$, denotes the following expression 
\beq
{\mathcal P}=\big(\EXP^{2\ii\beta}\,{\mathcal A}_1
+\EXP^{-2\ii\beta}\,{\mathcal D}_1\big)
\left(\EXP^{2\ii\gamma}\,
{\mathcal B}_1^*+\EXP^{-2\ii\gamma}\,{\mathcal C}_1^*\right)+
\big(\EXP^{-2\ii\beta}\,{\mathcal A}_1^*
+\EXP^{2\ii\beta}\,{\mathcal D}_1^*\big)
\big(\EXP^{-2\ii\gamma}\,
{\mathcal B}_1+\EXP^{2\ii\gamma}\,{\mathcal C}_1\big),
\eeq
which depends on spins and rapidity variables. A few more terms of the
expansion is given in Appendix~\ref{appB}. They are all manifestly
real and thereby confirm the validity of the star-star relation
\eqref{star-star} in perturbation
theory. Note that the leading term in \eqref{Vexp} is  
strictly positive. The
coefficients of the expansions in \eqref{WSexp} are analytic functions
of the spin variables, thus the integrals over central spins in the
star weight \eqref{weight-IRF} are non-singular in every order of
the expansion. 
It follows then that the coefficients in front of powers of $\pp$ and
$\qq$ in the RHS of
\eqref{Vexp} are analytic function of the external spins. Therefore, 
there exists a finite domain of the parameters $\pp$ and $\qq$ in the
vicinity of $\pp=\qq=0$, where the RHS of \eqref{Vexp} 
is strictly positive for all
values of the external spins $\boldsymbol{a} , \boldsymbol{b},
\boldsymbol{c}, \boldsymbol{d}$. We will call this domain a {\em
  physical regime}.
It would certainly be interesting to
investigate exact boundaries of this domain.       

In the physical regime the considered lattice model becomes 
a well defined model of statistical mechanics and Euclidean lattice
field theory.  Using the standard arguments based on the commutativity
of transfer matrices and functional equations for the
partition function per edge \cite{Str79,Zam79,Bax82inv}, we deduced that 
for a large number of sites $N$ (in the thermodynamic
limit) the bulk free energy of the model vanishes,
\begin{equation}
\lim_{N \to\infty} {N}^{-1}\,\log Z=0\;,\label{f-bulk}
\eeq
provided that the normalization \eqref{pf-edge} in \eqref{weights} is
taken into account.
  
\section{Quasi-classical limit}\label{sec:quasi-classical}
\subsection{Asymptotics of the Boltzmann weights}
Let us now consider the quasi-classical limit of our model, when one
of the nomes is real and fixed, while the other one tends to unity, 
\begin{equation}
\qq=\EXP^{\ii\pi\tau}=\textrm{real},\qquad
\pp=\EXP^{i\pi\sigma}\to1,\qquad \sigma\to0\,.\label{qclas}
\end{equation}
It convenient to use a rescaled parameter 
\beq
\hbar=-\ii\pi\sigma\to0\,,\label{hbar}
\eeq 
which plays the role of the Planck constant from the point of view of
Euclidean field theory (or the temperature from the point
of view of classical statistical mechanics). 
Introduce a new function
\begin{equation}
\lambda_4(z\,|\,\tau)\;=\;-\ii\int_0^z dw \log 
\overline\vartheta_4(z\,|\,\tau)\;,\label{lam-def}
\end{equation}
where $\overline\vartheta_j(z\,|\,\tau)$ is defined in
\eqref{thetabar}.
In the limit
\eqref{qclas} the elliptic
gamma-function \eqref{Phi-def} and the factor \eqref{pf-edge} become
\begin{equation}
\log \Phi(z)\;=\;-\frac{1}
{\hbar}\,\lambda_4(z\,|\,\tau)/\hbar+\mathcal{O}(\hbar^0)\,,\qquad 
\log \kappa_n(\alpha)=-\frac{1}
{\hbar}\,\lambda_4(\ii n \alpha\,|\,n \tau)+\mathcal{O}(\hbar^0),\label{Phi-as}
\eeq
moreover 
\beq
\log G(\sigma)=-\frac{\pi^2}{12\hbar}-\hf\log\hbar \,
+\mathcal{O}(\hbar^0),\quad 
\log \overline\vartheta_1(x\,|\,\sigma)=
-\frac{1}{\hbar}\Big\{x^2-\pi x+\frac{\pi^2}{6}\Big\}
+\mathcal{O}(\hbar^0),\quad 0<x<{\pi}\,,\label{theta-as}
\eeq
where $G(\sigma)$ is defined in \eqref{thetabar}. 
In writing quasi-classical expansions of the Boltzmann weights 
it is convenient to employ the permutation symmetry \eqref{ws-sym}.
Below we will assume that components of spin variables are always
arranged such that 
\begin{equation}\label{arrange}
\boldsymbol{x}=(x_1,x_2,\dots,x_n)\,,
\qquad -\frac{\pi}{2}\le\textrm{Re}(x_1)
<\textrm{Re}(x_2)<\cdots<\textrm{Re}(x_n)<\frac{\pi}{2}\,.
\end{equation}
 The leading quasi-classical asymptotics of the weights 
\eqref{weights} reads 
\bea
\log\iW_\alpha(\boldsymbol{x},\boldsymbol{y})&=&  -\frac{1}{\hbar}\,
\Lambda_\alpha(\boldsymbol{x},\boldsymbol{y})+{\mathcal
  O}(\hbar^0)\,,\\[5mm]
\log \overline{\iW}_\alpha(\boldsymbol{x},\boldsymbol{y})& =& 
-\frac{1}{\hbar}\,\overline{\Lambda}_\alpha(\boldsymbol{x},\boldsymbol{y})
-\frac{(n-1)}{2}\log\hbar +{\mathcal   O}(\hbar^0)\,, 
 \label{weight-exp}
\eea
where 
\begin{eqnarray}
\Lambda_\alpha(\boldsymbol{x},\boldsymbol{y})&=&\ds-\lambda_4(\ii n
\alpha\,|\,n\tau) +
\sum_{j,k} \lambda_4(x_j-y_k+\ii\alpha\,|\,\tau)\,,
\label{Lam-def}\\[8mm]
\overline{\Lambda}_\alpha(\boldsymbol{x},\boldsymbol{y})&=&
\Lambda_{\eta_0-\alpha}(\boldsymbol{x},\boldsymbol{y})
+ \frac{1}{2}{\mathcal C}(\boldsymbol{x})+\frac{1}{2}
{\mathcal C}(\boldsymbol{y})\,, \qquad \eta_0=-\ii\pi\tau/2. \label{Lambar1}
\end{eqnarray}
Eqs.\eqref{Lam-def} and \eqref{Lambar1} are trivial corollaries of
the definition \eqref{weights} and the expansions \eqref{Phi-as}.
The function ${\mathcal C}(\boldsymbol{x})$ is determined by the
asymptotics of the single-spin weight 
\beq
\log\iS(\boldsymbol{x})=
-\frac{1}{\hbar}\,{\mathcal
  C}(\boldsymbol{x})-\frac{(n-1)}{2}\log\hbar \,
+{\mathcal
  O}(\hbar^0)\,,
\eeq
defined in \eqref{S-def}. Using \eqref{theta-as} one obtains 
\beq
{\mathcal C}(\boldsymbol{x})=
\frac{(n^2-1)\,\pi^2}{12}
+n\sum_{j=1}^n x_j^2-2\pi\sum_{j=1}^n j\,x_j\,.\label{C-as}
\eeq
provided the variables $x_j$ are arranged as in \eqref{arrange}.
Finally, taking into account that $\qq$ is assumed real, one deduces 
that under the complex conjugation 
\beq
\Big({\Lambda}_\alpha(\boldsymbol{x},\boldsymbol{y})\Big)^*=
{\Lambda}_{\alpha^*}(\boldsymbol{y}^*,\boldsymbol{x}^*)\,, \label{lam-conj}
\eeq
where $\boldsymbol{x}^*=(x_1^*,x_2^*,\ldots,x_n^*)$ and similarly for
$\boldsymbol{y}^*$. 
\subsection{Classical star-star relation}
Consider now the quasi-classical limit for the composite 
Boltzmann weights of the
stars shown in Fig.~\ref{fig-IRF}. To make equations more compact
we will write the spin arguments
$\boldsymbol{a},\boldsymbol{b},\boldsymbol{c},\boldsymbol{d}$ 
\ on the same line, 
but assume the same spacial arrangement as in Fig.~\ref{fig-IRF}.
Substituting \eqref{weight-exp} into \eqref{V1} and 
calculating the integral by the saddle point method, one gets
\beq
\log\,\Big(\mathbb{V}_{\boldsymbol{u}\boldsymbol{v}}^{(1)}(
\boldsymbol{a},\boldsymbol{b},\boldsymbol{c},\boldsymbol{d})\Big)
=-\frac{1}{\hbar}\,
\mathcal{L}^{(1)}_{\boldsymbol{u},\boldsymbol{v}}
(\boldsymbol{X}\,
|\, \boldsymbol{a},\boldsymbol{b},\boldsymbol{c},\boldsymbol{d})
-\frac{(n-1)}{2}\log\hbar+
{\mathcal O}(\hbar^0)\,,
\eeq
where 
\beq
\mathcal{L}^{(1)}_{\boldsymbol{u},\boldsymbol{v}}\,
(\boldsymbol{x}\,
|\,\boldsymbol{a},\boldsymbol{b},\boldsymbol{c},\boldsymbol{d})=  
\;\overline{\Lambda}_{u-v}(\boldsymbol{c},\boldsymbol{x})+
\overline{\Lambda}_{u'-v'}(\boldsymbol{b},\boldsymbol{x})+
\Lambda_{u'-v}(\boldsymbol{x},\boldsymbol{a})+
\Lambda_{u-v'}(\boldsymbol{x},\boldsymbol{d})\label{action1}
\eeq
and $\boldsymbol{X}=(X_1,X_2,\ldots,X_n)$\  
is the saddle point, 
determined by the equations 
\beq
\Big(\frac{\partial}{\partial x_k}-\frac{\partial}{\partial
  x_{k+1}}\Big) \ \mathcal{L}^{(1)}_{\boldsymbol{u},\boldsymbol{v}}
(\boldsymbol{x}\,|\,\boldsymbol{a},\boldsymbol{b},\boldsymbol{c},
\boldsymbol{d}) 
\,\Big\vert_{\boldsymbol{x}=\boldsymbol{X}}=0,\qquad 
k=1,2,\ldots,n-1\,.\label{sad-point}
\eeq
Remind that $\boldsymbol{x}$ is the 
multi-component variable,
$\boldsymbol{x}=(x_1,x_2,\ldots,x_n)$, whose components $x_k$ are
constrained by \eqref{spinvar} and cannot be varied independently. 
That is why there are only $(n-1)$
equations in  \eqref{sad-point}, each containing a difference of two
partial derivatives. 
Define new variables 
\beq\begin{array}{rclrclrcl}
\alpha_1&=&u'-v\,,\qquad&\alpha_2&=&\eta_0-u'+v'\,,\qquad
&\alpha_3&=&\eta_0-u+v\,,\\[3mm]
\alpha_4&=&u-v'\,,\qquad&
\alpha_5&=& u'-u\,,\qquad&\alpha_6&=& v'-v\,,
\end{array}\label{alphas}
\eeq
constrained by three relation 
\beq
\alpha_1+\alpha_2+\alpha_3+\alpha_4=2\eta_0\,,\qquad
\alpha_5=\alpha_1+\alpha_3-\eta_0\,,\qquad
\alpha_6=\alpha_1+\alpha_2-\eta_0\,.\label{const}
\eeq
Below we will regard $\alpha_1,\alpha_2,\alpha_3$ 
as independent variables. 
Introduce a new function 
\beq\begin{array}{rcl}
\psi(X\,|\,\boldsymbol{a},\boldsymbol{b},\boldsymbol{c},
\boldsymbol{d}\,|\,\alpha_1,\alpha_2,\alpha_3,\alpha_4) &=\ds-\ii\,
\sum_{k=1}^n \log\frac{\vartheta_4(X-a_k+\ii\alpha_1\,|\,\tau)\,
\vartheta_4(X-d_k+\ii\alpha_4\,|\,\tau)}
{\vartheta_4(c_k-X+\ii\alpha_3\,|\,\tau)\,
\vartheta_4(b_k-X+\ii \alpha_2\,|\,\tau)}\;,
\end{array}\label{gg-def}
\end{equation}
which depend on a (single-component) variable $X$ and four 
multi-component spin variables $
\boldsymbol{a},\boldsymbol{b},\boldsymbol{c},
\boldsymbol{d}$ and the rapidity-type variables 
$\alpha_1,\alpha_2,\alpha_3,\alpha_4$ (it is also implicitly depends
on the elliptic modular parameter $\tau$). 
Using \eqref{Lam-def}, \eqref{Lambar1} and \eqref{C-as} one can
write the variational equations \eqref{sad-point} in the form 
\beq\begin{array}{l}
\psi(X_{k+1}\,|\,\boldsymbol{a},\boldsymbol{b},\boldsymbol{c},
\boldsymbol{d}\,|\,\alpha_1,\alpha_2,\alpha_3,\alpha_4)
-
\psi(X_{k}\,|\,\boldsymbol{a},\boldsymbol{b},\boldsymbol{c},
\boldsymbol{d}\,|\,\alpha_1,\alpha_2,\alpha_3,\alpha_4)=\\[5mm]
\phantom{XXXXXXXXXXXXXX}
=+2\pi-2n\,(X_{k+1}-X_k)
\,, 
\qquad k=1,2,\ldots,n-1\,,
\end{array}
\label{vareq1}
\eeq
where the RHS comes from a variation of the function ${\mathcal
  C}(\boldsymbol{x})$ associated with the central spin in the star (it
enters the expression \eqref{Lambar1}).
Similarly, for the second 
star in Fig.~\ref{fig-IRF}, one gets
\beq
\log\,\Big(\mathbb{V}_{\boldsymbol{u}\boldsymbol{v}}^{(2)}(
\boldsymbol{a},\boldsymbol{b},\boldsymbol{c},\boldsymbol{d})\Big)
=-\frac{1}{\hbar}\,
\mathcal{L}^{(2)}_{\boldsymbol{u},\boldsymbol{v}}
(\boldsymbol{Y}\,
|\, \boldsymbol{a},\boldsymbol{b},\boldsymbol{c},\boldsymbol{d})
-\frac{(n-1)}{2}\log\hbar+
{\mathcal
  O}(\hbar^0)\,,
\eeq
where 
\beq
\mathcal{L}^{(2)}_{\boldsymbol{u},\boldsymbol{v}}\,
(\boldsymbol{y}\,
|\,\boldsymbol{a},\boldsymbol{b},\boldsymbol{c},\boldsymbol{d})\;=  
\;\overline{\Lambda}_{u-v}(\boldsymbol{y},\boldsymbol{b})+
\overline{\Lambda}_{u'-v'}(\boldsymbol{y},\boldsymbol{c})+
\Lambda_{u'-v}(\boldsymbol{d},\boldsymbol{y})+
\Lambda_{u-v'}(\boldsymbol{a},\boldsymbol{y})\,,\label{action2}
\eeq
and the saddle point $\boldsymbol{Y}=(Y_1,Y_2,\ldots,Y_n)$\  
is determined by the variational equations\footnote{%
Notice different signs in the RHS of this equation with
respect to \eqref{vareq1}.}  
\beq\begin{array}{l}
\psi(Y_{k+1}\,|\,\boldsymbol{a},\boldsymbol{b},\boldsymbol{c},
\boldsymbol{d}\,|\, {-\alpha_4},-\alpha_3,-\alpha_2,-\alpha_1)
-
\psi(Y_{k}\,|\,\boldsymbol{a},\boldsymbol{b},\boldsymbol{c},
\boldsymbol{d}\,|\,  {-\alpha_4},-\alpha_3,-\alpha_2,-\alpha_1)=\\[5mm]
\phantom{XXXXXXXXXXXXXX}
=-2\pi+2n\,(Y_{k+1}-Y_k)
\,, 
\qquad k=1,2,\ldots,n-1\,.
\end{array}
\label{vareq2}
\eeq

The quasi-classical expansion of the {\em quantum}
star-star relation \eqref{star-star} leads an infinite
number of non-trivial relations --- one relation for each order of
the expansion in $\hbar$. In the leading order one obtains,
\beq
\mathcal{L}^{(1)}_{\boldsymbol{u},\boldsymbol{v}}\,
(\boldsymbol{X}
|\boldsymbol{a},\boldsymbol{b},\boldsymbol{c},\boldsymbol{d})-
\Delta_{\boldsymbol{u},\boldsymbol{v}}
(\boldsymbol{a},\boldsymbol{b},\boldsymbol{c},\boldsymbol{d})=  
\mathcal{L}^{(2)}_{\boldsymbol{u},\boldsymbol{v}}\,
(\boldsymbol{Y}
|\boldsymbol{a},\boldsymbol{b},\boldsymbol{c},\boldsymbol{d})
+
\Delta_{\boldsymbol{u},\boldsymbol{v}}
(\boldsymbol{a},\boldsymbol{b},\boldsymbol{c},\boldsymbol{d})\,,\label{cstar}
\eeq
where $\boldsymbol{X}$ and $\boldsymbol{Y}$ are determined by 
\eqref{vareq1} and \eqref{vareq2} and
\begin{equation}
\Delta_{\boldsymbol{u},\boldsymbol{v}}
(\boldsymbol{a},\boldsymbol{b},\boldsymbol{c},\boldsymbol{d})=  
\frac{1}{2}\left(\Lambda_{v'-v}(\boldsymbol{b},\boldsymbol{a})
+\Lambda_{u'-u}(\boldsymbol{c},\boldsymbol{a})
-\Lambda_{v'-v}(\boldsymbol{d},\boldsymbol{c})
-\Lambda_{u'-u}(\boldsymbol{d},\boldsymbol{b})\right)\;.\label{delta-def}
\end{equation}
We name the formula \eqref{cstar} a {\em classical star-star relation}. 

For the weight function \eqref{weight-IRF} one obtains
\beq
\log\mathbb{V}_{\boldsymbol{u}\boldsymbol{v}}(\boldsymbol{a},\boldsymbol{b},
\boldsymbol{c}, \boldsymbol{d})=-\frac{1}{\hbar} 
\mathcal{L}_{\boldsymbol{u},\boldsymbol{v}}(\boldsymbol{a},\boldsymbol{b},\boldsymbol{c},\boldsymbol{d})
-\frac{(n-1)}{2}\log\hbar+{\mathcal O}(\hbar)\,,\label{VV-as}
\eeq
where the Lagrangian density $\mathcal{L}_{\boldsymbol{u},
\boldsymbol{v}}(\boldsymbol{a},\boldsymbol{b},\boldsymbol{c},\boldsymbol{d})$ 
coincides with the LHS (or the RHS) of the \eqref{cstar}. Writing it
in full, one obtains 
\begin{equation}
\begin{array}{rcl}
\mathcal{L}_{\boldsymbol{u},\boldsymbol{v}}(\boldsymbol{a},\boldsymbol{b},\boldsymbol{c},\boldsymbol{d})
&=&\overline{\Lambda}_{u-v}(\boldsymbol{c},\boldsymbol{X})+
\overline{\Lambda}_{u'-v'}(\boldsymbol{b},\boldsymbol{X})+
\Lambda_{u'-v}(\boldsymbol{X},\boldsymbol{a})+
\Lambda_{u-v'}(\boldsymbol{X},\boldsymbol{d})-
\Delta_{\boldsymbol{u},\boldsymbol{v}}
(\boldsymbol{a},\boldsymbol{b},\boldsymbol{c},\boldsymbol{d})\\
[5mm]
&=&\overline{\Lambda}_{u-v}(\boldsymbol{Y},\boldsymbol{b})+
\overline{\Lambda}_{u'-v'}(\boldsymbol{Y},\boldsymbol{c})+
\Lambda_{u'-v}(\boldsymbol{d},\boldsymbol{Y})+
\Lambda_{u-v'}(\boldsymbol{a},\boldsymbol{Y})+
\Delta_{\boldsymbol{u},\boldsymbol{v}}
(\boldsymbol{a},\boldsymbol{b},\boldsymbol{c},\boldsymbol{d})\,,
\end{array}\label{lagrange}
\end{equation}
where $\boldsymbol{X}$ and $\boldsymbol{Y}$ are determined by
the variational equations \eqref{vareq1} and \eqref{vareq2}. 

Consider
now the regime \eqref{pq-rel} with real $\qq$ as in
\eqref{qclas}. Assume that the spins $
\boldsymbol{a},\boldsymbol{b},\boldsymbol{c},\boldsymbol{d}$ are real.
Then one can show that if $\boldsymbol{X}$ solves  \eqref{vareq1}
then $\boldsymbol{Y}=\boldsymbol{X}^*$ solves \eqref{vareq2}, where
the star denotes the complex conjugation. Having this in mind and
using \eqref{lam-conj} it is easy to check that the complex conjugation just
interchanges two alternative expressions in \eqref{lagrange}. This
means that the Lagrangian density $\mathcal{L}_{\boldsymbol{u},
\boldsymbol{v}}(\boldsymbol{a},\boldsymbol{b},\boldsymbol{c},\boldsymbol{d})$ 
is real. 

From \eqref{weights} and \eqref{weight-exp} it follows that 
\beq
\Lambda_\alpha(\boldsymbol{a} ,\boldsymbol{b}) +
\Lambda_{-\alpha}(\boldsymbol{b} ,\boldsymbol{a})=0\,. 
\eeq
Using this equality together with \eqref{Lambar1}, substituting
\eqref{action1}, \eqref{action2} and \eqref{delta-def} into
the classical star-star relation \eqref{cstar} and moving all 
terms there to one side, one obtains
\beq
\begin{array}{l}
\phantom{+}\,\Lambda_{\alpha_1}(\boldsymbol{X},\boldsymbol{a})+
\Lambda_{\alpha_2}(\boldsymbol{b},\boldsymbol{X})+
\Lambda_{\alpha_3}(\boldsymbol{c},\boldsymbol{X})+
\Lambda_{\alpha_4}(\boldsymbol{X},\boldsymbol{d})
+\,\Lambda_{\alpha_6}(\boldsymbol{d},\boldsymbol{c})+
\Lambda_{\alpha_5}(\boldsymbol{d},\boldsymbol{b})\\[4mm]
+\,\Lambda_{-\alpha_4}(\boldsymbol{Y},\boldsymbol{a})+
\Lambda_{-\alpha_3}(\boldsymbol{b},\boldsymbol{Y})+
\Lambda_{-\alpha_2}(\boldsymbol{c},\boldsymbol{Y})+
\Lambda_{-\alpha_1}(\boldsymbol{Y},\boldsymbol{d})
+\Lambda_{-\alpha_6}(\boldsymbol{a},\boldsymbol{b})+
\Lambda_{-\alpha_5}(\boldsymbol{a},\boldsymbol{c})\\[4mm]
+\,\mathcal{C}(\boldsymbol{X})\,-\,
\mathcal{C}(\boldsymbol{Y})=0\,.
\end{array}\label{cstar1}
\eeq
The terms of this relation can be conveniently associated 
\begin{figure}{h}
\centering
\includegraphics[scale=.65]{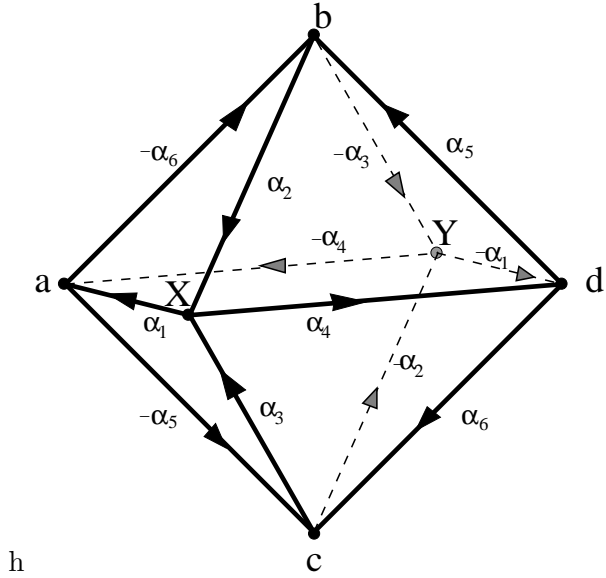}
\caption{Graphical representation of the classical star-star relation
  \eqref{cstar1}.}  \label{octa}
\end{figure}
with edges and vertices of an octahedron. 
Consider a perfect octahedron shown in Fig.~\ref{octa}. 
Its edges are
oriented, as indicated by arrows, such that there are exactly two incoming
and two outgoing edges at each vertex. The vertices are assigned to the
spin variables $\boldsymbol{a},\boldsymbol{b},\boldsymbol{c},
\boldsymbol{d},\boldsymbol{X},\boldsymbol{Y}$, as indicated. 
The edges carry rapidity-type variables
$\pm\alpha_1,\pm\alpha_2,\ldots,\pm\alpha_6$, 
shown near edges.  A perfect octahedron contains six pairs of
parallel edges. The values of the $\alpha$-variables on parallel edges
are equal in absolute value, but differ in signs. 
Moreover, the $\alpha$-variables are constrained by the relations
\eqref{const}. Each edge
corresponds to a $\Lambda$-term in \eqref{cstar1}.
For example, the edge
directed from ${\boldsymbol{X}}$ to $\boldsymbol{a}$ and carrying the rapidity
label $\alpha_1$ in Fig.~\ref{octa} corresponds to the term
$\Lambda_{\alpha_1}(\boldsymbol{X},\boldsymbol{a})$ in
\eqref{cstar1}. Similarly for all other edges. 
Further, according 
to \eqref{const}, the arithmetic sum of the $\alpha$-variables 
on four edges meeting at the vertex $\boldsymbol{X}$ (regardless their
directions) is equal to
$+2\pi$. This vertex contributes the term 
``$+\mathcal{C}(\boldsymbol{X})$'' to
\eqref{cstar1}. Similarly, the sum of $\alpha$'s at the vertex 
$\boldsymbol{Y}$ is equal to $-2\pi$. This vertex contributes the term
``$-\mathcal{C}(\boldsymbol{Y})$''. Finally, thanks to \eqref{const}
the sum of $\alpha$'s
for each of the remaining four vertices $\boldsymbol{a},\boldsymbol{b},
\boldsymbol{c}, \boldsymbol{d}$ exactly vanishes. These vertices do not
contribute any ${\mathcal C}$-terms into \eqref{cstar1}.
        
From mathematical point of view it is convenient to regard equations
\eqref{vareq1} and \eqref{vareq2}  
as constrains placed on six spin variables at the vertices of the octahedron.
Interestingly, these constrains can be re-written in two other
equivalent forms. Differentiating \eqref{cstar1} with respect to the
spins $\boldsymbol{a}$ and $\boldsymbol{d}$ one obtains
\beq
\left\{\begin{array}{l}
\psi(a_{k+1}\,|\,\boldsymbol{b},\boldsymbol{X},\boldsymbol{Y},
\boldsymbol{c}\,|\, {-\alpha_6},+\alpha_1,-\alpha_4,-\alpha_5)
-
\psi(a_{k}\,|\,\boldsymbol{b},\boldsymbol{X},\boldsymbol{Y},
\boldsymbol{c}\,|\, {-\alpha_6},+\alpha_1,-\alpha_4,-\alpha_5)=0
\,, \\[5mm]
\psi(d_{k+1}\,|\,\boldsymbol{b},\boldsymbol{X},\boldsymbol{Y},
\boldsymbol{c}\,|\,  {+\alpha_5},+\alpha_4,-\alpha_1,+\alpha_6)
-
\psi(d_{k}\,|\,\boldsymbol{b},\boldsymbol{X},\boldsymbol{Y},
\boldsymbol{c}\,|\,  {+\alpha_5},+\alpha_4,-\alpha_1,+\alpha_6)=0
\,, 
\end{array}\right.
\label{vareq_new1}
\eeq
where $ k=1,2,\ldots,n-1$. Here we have used the fact that the
LHS of \eqref{cstar1} is stationary with respect to variations of
$\boldsymbol{X}$ and $\boldsymbol{Y}$ by virtue 
of \eqref{vareq1} and \eqref{vareq2}.  Similarly, 
differentiating \eqref{cstar1} with respect to the
spins $\boldsymbol{b}$ and $\boldsymbol{c}$ one obtains
\beq
\left\{\begin{array}{l}
\psi(b_{k+1}\,|\,\boldsymbol{X},\boldsymbol{a},\boldsymbol{d},
\boldsymbol{Y}\,|\, {+\alpha_2},-\alpha_6,+\alpha_5,-\alpha_3)
-
\psi(b_{k}\,|\,\boldsymbol{X},\boldsymbol{a},\boldsymbol{d},
\boldsymbol{Y}\,|\, {+\alpha_2},-\alpha_6,+\alpha_5,-\alpha_3)=0
\,, \\[5mm]
\psi(c_{k+1}\,|\,\boldsymbol{X},\boldsymbol{a},\boldsymbol{d},
\boldsymbol{Y}\,|\,{+\alpha_3},-\alpha_5,+\alpha_6,-\alpha_2)
-
\psi(c_{k}\,|\,\boldsymbol{X},\boldsymbol{a},\boldsymbol{d},
\boldsymbol{Y}\,|\,{+\alpha_3},-\alpha_5,+\alpha_6,-\alpha_2)=0
\,, 
\end{array}\right.
\label{vareq_new2}
\eeq
where $ k=1,2,\ldots,n-1$. 

\subsection{Discrete non-linear equations}

In the limit \eqref{hbar} the partition function \eqref{Z-IRF} 
develops a typical quasi-classical asymptotics. Substituting
\eqref{VV-as} into \eqref{Z-IRF} and calculating the integral by the
saddle point method one obtains,
\begin{equation}
\log Z = -\frac{1}{\hbar}
\mathcal{A}(\{\boldsymbol{y}\})+{\mathcal O}(\hbar^0)
\end{equation}
where the action
\beq
\mathcal{A}(\{\boldsymbol{y}\})=\sum_{\textrm{white stars}} 
\mathcal{L}_{\boldsymbol{u}, 
\boldsymbol{v}}(\boldsymbol{a},\boldsymbol{b},\boldsymbol{c},
\boldsymbol{d})\label{action-full} 
\eeq
where the sum over all white-centred stars and 
$\boldsymbol{a},\boldsymbol{b},\boldsymbol{c},\boldsymbol{d}$ denote
the corresponding outer spins arranged as in Fig.~\ref{fig-IRF}. Here
$\{\boldsymbol{y}\}$ denotes an equilibrium configuration of spins on
the black sub-lattice, which is determined by the Euler-Lagrange equations 
\begin{equation}\label{Euler}
\frac{\delta}{\delta \boldsymbol{y}(\vec{r})} \,
\mathcal{A}(\{\boldsymbol{y}\})\;=\;0\;,
\end{equation}
where $\vec{r}$ is an integer two-dimensional vector numerating 
the internal sites on the black sub-lattice 
(remind that the spins are varied subject
to the constraints \eqref{spinvar}).  
Using the first expression from \eqref{lagrange} one could bring these
equations to the form 
\begin{equation}\label{Euler-b}
\frac{\delta}{\delta \boldsymbol{y}}\, 
\mathcal{L}^{(2)}_{\boldsymbol{u},\boldsymbol{v}}\,
(\boldsymbol{y}\,
|\,\boldsymbol{a},\boldsymbol{b},\boldsymbol{c},\boldsymbol{d})\;=0  
\end{equation}
for every black site. The function 
$\mathcal{L}^{(2)}_{\boldsymbol{u},\boldsymbol{v}}\,
(\boldsymbol{y}\,
|\,\boldsymbol{a},\boldsymbol{b},\boldsymbol{c},\boldsymbol{d})$
is defined in \eqref{action2} and the symbols 
$\boldsymbol{y}, \boldsymbol{a},\boldsymbol{b},\boldsymbol{c},\boldsymbol{d}
$ denote the spins on a (black-centred) star, arranged as in
Fig.~\ref{fig-IRF} on the right side. The equations \eqref{Euler-b}
are explicitly presented in \eqref{vareq2} in an expanded form. 
Further, the definition 
\eqref{lagrange} for the Lagrangian $
\mathcal{L}_{\boldsymbol{u},
\boldsymbol{v}}(\boldsymbol{a},\boldsymbol{b},\boldsymbol{c},\boldsymbol{d})$, 
entering the action \eqref{action-full}, involves the variational equations
\eqref{sad-point}, 
\begin{equation}\label{Euler-a}
\frac{\delta}{\delta \boldsymbol{x}}\, 
\mathcal{L}^{(1)}_{\boldsymbol{u},\boldsymbol{v}}\,
(\boldsymbol{x}\,
|\,\boldsymbol{a},\boldsymbol{b},\boldsymbol{c},\boldsymbol{d})\;=0  
\end{equation}
for every white site. These equations are explicitly presented 
in \eqref{vareq1}. Combining the equations \eqref{Euler-b} and
\eqref{Euler-a} (respectively, their expanded forms
\eqref{vareq1}, \eqref{vareq2}), one gets a system of non-linear
difference equations for all internal sites of the lattice (the
boundary spins are fixed).  
Note, that each of these equations
involves variables on five sites, belonging to a four-edge star.  
These classical integrable equations can be
regarded as a generalization of the Laplace-type system
associated with the 
Adler-Bobenko-Suris $Q_4$ equation \cite{AS04,Bazhanov:2010kz} to the
case of  
multi-component field variables. Their integrability is inherited from
the integrability of the quantum model. It could also be independently
established from the classical star-star relation \eqref{cstar} (in
particular, the latter implies the existence of an infinite set of
local integrals of motion).

It appears that for Dirichlet (fixed) boundary conditions solutions of the
classical equation  \eqref{vareq1}, \eqref{vareq2} possess the
following property. We expect that in the bulk of a
large lattice (i.e. away from the boundary) all solutions 
converge to some constant solution, independent of 
the boundary values of the spins. 
According to \eqref{f-bulk} it is reasonable to expect
that the action \eqref{action-full} evaluated on this constant
solution should vanish. The corresponding solution could be easily
found. Assume that all spins on the
lattice are equal to same vector
\beq
{\boldsymbol{x}}^{(c)}=(x_1,x_2,\ldots,x_n),\qquad 
x_j=\frac{\pi}{n}\Big(j-\frac{n+1}{2}\Big),\qquad j=1,2,\ldots,n\,.
\label{consol}
\eeq
Then it is easy to check that all equations  \eqref{vareq1}, \eqref{vareq2}
are satisfied and that 
\beq
\Lambda_\alpha(\boldsymbol{x}^{(c)},\boldsymbol{x}^{(c)})=
{\mathcal C}(\boldsymbol{x}^{(c)})=0\,.
\eeq
Therefore the action \eqref{action-full} exactly vanishes on this
solution in complete agreement with \eqref{f-bulk}. 
It would be interesting to verify whether this it is an
absolute minimum  of the action and, more generally, to study its
global convexity properties.  

As an illustration consider the case $n=2$. 
The two-component spin variables in this case 
contain 
only one independent continuous variable, 
so that it will be more convenient to use the scalar argument $x$,
assuming that the corresponding two-component spin is equal to 
$\boldsymbol{x}=(-x,x)$, and similarly for all other spins.
Eq. \eqref{Lam-def} and \eqref{C-as} simplify to 
\beq
\Lambda_\alpha(x,y)=-\ii \int_0^{x-y} dw \log \frac
       {\vartheta_4(w+\ii\alpha\,|\,\tau)}
{\vartheta_4(w-\ii\alpha\,|\,\tau)}  
-\ii \int_{\pi/2}^{x+y} dw \log \frac
       {\vartheta_4(w+\ii\alpha\,|\,\tau)}
{\vartheta_4(w-\ii\alpha\,|\,\tau)}\,, 
\eeq
and
\beq
{\mathcal C}(x)=\frac{\pi^2}{4}+4x^2-2\pi|x|\,,\qquad
-\frac{\pi}{2}\le x\le
\frac{\pi}{2}\,. \label{n2C}
\eeq
Note that for $n=2$ the function $\Lambda_\alpha(x,y)$  becomes
symmetric upon interchanging $x$ and $y$.
Using the above expressions in \eqref{Lambar1} and
\eqref{lagrange} one can bring the action \eqref{action-full} to the
form 
\beq
{\mathcal A}(\{x_s\})=\sum_{\langle rs \rangle}\Lambda_{\alpha_{rs}}(x_r,x_s) +
\sum_{s} {\mathcal C}(x_s)\label{n2action}
\eeq
where the first sum is over all edges ${\langle rs \rangle}$, while
the second sum is over all sites $s$. The variables $\alpha_{rs}$ take
one of the four values $\alpha_1,\alpha_2,\alpha_3,\alpha_4$, defined
in \eqref{alphas}, depending on the type of the edge ${\langle rs
  \rangle}$
and the values of
rapidity variables passing through the edge. The variational equations 
\eqref{Euler-a} and \eqref{vareq1} for the white-centred stars, as in
Fig.~\ref{fig-IRF}, reduce
to
\beq
\varphi_{\alpha_1}(x,a)+
\varphi_{\alpha_2}(x,b)+
\varphi_{\alpha_3}(x,c)+
\varphi_{\alpha_4}(x,d)=2\pi-8 x
\eeq
where 
\beq
\varphi_{\alpha}(x,y)=\partial_x \Lambda_a(x,y)=\frac{1}{\ii}\log
\frac{\vartheta_4(x-y+\ii \alpha)\vartheta_4(x+y+\ii \alpha)}
{\vartheta_4(x-y-\ii \alpha)\vartheta_4(x+y-\ii \alpha)}
\eeq
Similarly, for the black-centred stars one gets 
\beq
\varphi_{\alpha_4}(y,a)+
\varphi_{\alpha_3}(y,b)+
\varphi_{\alpha_2}(y,c)+
\varphi_{\alpha_1}(y,d)=2\pi-8 y
\eeq
where the spins are arranged as in Fig.~\ref{fig-IRF}. The 
constant solution \eqref{consol} in this case reads  
\beq
x^{(c)}=\frac{\pi}{4}, \label{iso}
\eeq
for all sites of the lattice. Observing that 
\beq
\Lambda_\alpha\Big(\frac{\pi}{4},\frac{\pi}{4}\Big)=
{\mathcal C}\Big(\frac{\pi}{4}\Big)=0\ ,
\eeq
one easily concludes that action \eqref{n2action} 
on this solution vanishes exactly
${\mathcal A}(\{x^{(c)}_j\})=0$. Let us show that this is the minimum
of the action\footnote{%
Since the expression \eqref{n2C} is non-analytic for $x=0$, one needs
to check the case where all spins are vanishing $x_s=0$. Careful
considerations show that the value of the action for this configuration is
greater than zero.}.
Expanding the latter in the vicinity of this solution 
\beq
x_s=\frac{\pi}{4}+\varepsilon_s,\qquad \varepsilon_s\to 0\,,
\eeq
one gets
\beq
{\mathcal A}(\{x_s\})=\ds\sum_{\langle rs
  \rangle}\Big(\zeta_4(\ii\alpha_{rs})-\zeta_3(\ii\alpha_{rs})\Big)\,
(\varepsilon_r-
\varepsilon_s)^2\,
+\ds2 c\sum_s \, \varepsilon_s^2+{\mathcal O}(\varepsilon^3),\label{qform}
\eeq
where 
\beq
\zeta_k(z)=\frac{1}{\ii}\,\frac{\partial }{\partial z}\log
\vartheta_k(z\,|\,\tau)\,, \qquad k=3,4,
\eeq
and 
\beq
c=2+\zeta_3(\ii\alpha_1)+
\zeta_3(\ii\alpha_2)+\zeta_3(\ii\alpha_3)+\zeta_3(\ii\alpha_4).
\eeq
Remind that $\tau$ is assumed to be purely imaginary, ${\rm
  Im}\,\tau>0$. Using the inequality
\beq
\zeta_4(\ii\alpha)>0>\zeta_3(\ii\alpha)>
-\frac{2\alpha}{\pi|\tau|}\,,\qquad 0<\alpha<\pi|\tau|/2
\eeq
and remembering that the sum of $\alpha$'s is constrained by
\eqref{const}, with $\eta_0=-\ii\pi\tau/2$, one can check 
that 
\beq
c>0, \qquad 0<\alpha_k<\eta_0, \qquad k=1,2,3,4, 
\eeq
and thus that \eqref{qform} defines a positive-definite quadratic
form. The constant solution \eqref{iso} could be though as an analog 
of the isoradial solution of the Hirota equations describing 
planar circle patterns \cite{BSp,BMS:2005}.

\section{Conclusion}

We formulated a new solvable edge-interaction model 
of statistical mechanics with multi-component continuous spin
variables taking values on a circle. The Boltzmann weights are given
by Eqs. \eqref{weights}, \eqref{S-def}. The weights satisfy 
the star-star relation \eqref{star-star}, which ensures the
integrability of the model. Currently, we claim this relation as a
conjecture. We have verified this relation in a few orders of
perturbation theory 
in the temperature-like variables (see Appendix~\ref{appB}) and made
extensive numerical checks for $n=3,4$. 
We expect that a complete proof could be obtained by a rather
straightforward generalization of the results of
\cite{Bazhanov:1990qk,Date:1990bs,Kashaev:1990wy,Bazhanov:1992jqa,Znbrok}, 
devoted to discrete-spin models connected with the $sl(n)$ algebra. 

The star-star relation \eqref{star-star} implies the validity of the
Yang-Baxter equations in the vertex \eqref{YBE-int} and IRF
\eqref{YBE-IRF} forms. The partition function per edge of the lattice
is calculated exactly in the thermodynamic limit by means of the
inversion relations \cite{Str79,Zam79,Bax82inv}. The result is
included into the normalization of the Boltzmann weights given by
\eqref{ks-def}, \eqref{pf-edge}. With this normalization the bulk free
energy of model vanishes, see \eqref{f-bulk}. 

In Section~\ref{sec:quasi-classical} we study the quasi-classical (or
the low-temperature) limit when one of the elliptic nomes tends to
unity, $\pp\to1$, and obtain the classical version (Eqs.\eqref{cstar}
and \eqref{cstar1}) of the star-star relation. 

Here we have only formulated our model on a simple square lattice, but
the latter can be replaced with a rather general planar graph. The
required construction is similar to that described in
\cite{Bazhanov:2010kz} with some modifications accommodating the two
types of rapidity lines shown in Fig.~\ref{fig-crosses}. The partition
function then will be invariant under certain deformations of the
planar graph connected with the star-star \eqref{star-star} and
inversion \eqref{inversion} relations. This property is known as
Baxter's $Z$-invariance \cite{Bax1}. Evidently, this property
continues to hold in quasi-classical limit. As a result the action of
the classical system \eqref{action-full} evaluated on the stationary
configuration will be invariant under the ``star-star''
transformations connected with the classical star-star relation
\eqref{cstar1}. These moves are similar to the ``star-triangular''
transformations for the classical action introduced in
\cite{BMS07a}. Remind that, geometrically, the classical star-triangle
relation can be associated with a tetrahedron formed by a set of six
face diagonals of a cube. By this reason the star-triangular moves are
often referred to as ``cubic flips'' of faces of a quadrilateral
surface \cite{BS09}. From this point of view the star-star relation
\eqref{cstar1} could be associated with ``rhombic-dodecahedral flips'' of
quadrilateral surfaces, since the octahedron in Fig.~\ref{octa},
representing the various terms in \eqref{cstar1}, forms a set of
face diagonals of a rhombic dodecahedron.

As mentioned in the introduction, for $n=2$ the model provides a
master solution \cite{Bazhanov:2010kz} of the Yang-Baxter equation,
which contains all known edge-interaction models with single-component
spins as particular cases. It turns out that for $n>2$ the model also
possesses a similar ``master'' property. The Boltzmann weights \eqref{weights}
allow a large number of interesting limiting cases. 
For example, one could consider a low-temperature limit
when one of the nomes tends to a root  
of unity,
\beq
\pp\to\EXP^{\ii\pi/N}\,,\qquad N=2,3,4,\ldots \,.
\eeq
This limit is similar, but more complicated than the quasi-classical
limit considered in Section~\ref{sec:quasi-classical}. 
In this case
one obtains a ``hybrid'' model which couples a classical integrable
system, involving continuous spin variables, and an Ising-type model 
of statistical mechanics with discrete multi-component spins
variable taking values in $(\mathbb{Z}_N)^{n-1}$. 
The Boltzmann weights are expressed with elliptic theta functions. 
In general, the emerging model is spatially inhomogeneous, since its 
Boltzmann weights depend on solutions of the classical equation of motion
\footnote{%
These classical equation of motion are very similar to 
Eqs.\eqref{Euler-a} and \eqref{Euler-b}. 
Conceptually the hybrid model described here 
is very much similar to a model of quantum
field theory on a non-trivial classical background.}. 
In a particular homogeneous case connected to a
constant classical solution (similar to 
\eqref{consol}) this model reduces
to the $sl_n$-generalization \cite{Znbrok} of the Kashiwara-Miwa model
\cite{Kashiwara:1986}. In the trigonometric limit this elliptic
model further reduces to the $sl_n$-generalized chiral Potts model
\cite{Bazhanov:1990qk,Date:1990bs}, which is in its turn 
equivalent to the $N$-state
Zamolodchikov-Bazhanov-Baxter model \cite{Zamolodchikov:1981kf,
Bazhanov:1992jqa} on a cubic lattice (the number $n$ becomes the size
of the lattice in the ``hidden'' third dimension). We hope to
consider all these limits and connections in the future (the $n=2$
case has been thoroughly studied in \cite{Bazhanov:2010kz}). It would
also be interesting to explore possible connections of our results to
the discrete-spin models considered in \cite{Zabrodin:2010}.

\section*{Acknowledgments} 
The authors thank R.J.~Baxter, A.I.~Bobenko, V.V.~Mangazeev,
 V.P.~Spiridonov and Y.B.~Suris for interesting discussions. This work
 was partially supported by the Australian Research Council.

\app{Inversion relation}\label{appA}
\noindent
Theorem 11 from \cite{Spiridonov-essays} (related to $A_{n-1}$ root system)  
can be re-written in our notations as
\begin{equation}
\begin{array}{l}
\ds \int d\boldsymbol{y}\,
\overline{\iW}_{\alpha}(\boldsymbol{x},\boldsymbol{y})\,
\overline{\iW}_\beta(\boldsymbol{y},\boldsymbol{z}) \,
\prod_{j=1}^n \Phi(\sigma-y_j+\ii n \alpha) \,\Phi(y_j-\sigma +\ii
n\beta)\;=
\\
[5mm]
\ds
\phantom{XX}
\;=\;\overline{\iW}_{\alpha+\beta}(\boldsymbol{x},\boldsymbol{z}) 
\,\frac{\kappa_n(\eta-\alpha-\beta)\,\Phi(\ii\eta-\ii n
  \alpha)\,\Phi(\ii\eta-\ii n \beta)}{\Phi(\ii\eta-\ii
  n(\alpha+\beta))\kappa_n(\eta-\alpha)\,\kappa_n(\eta-\beta)}\\ 
[5mm]
\phantom{XX}\ds\times \prod_{j=1}^n \Phi(\sigma-x_j+\ii\alpha)\,
\Phi(z_j-\sigma+\ii\beta)\, 
\Phi(x_j-\sigma +\ii(n\beta-\alpha))\, \Phi(\sigma-z_j + \ii(n\alpha-\beta))\;.
\end{array}
\end{equation}
Taking the limit $\beta\to -\alpha$ and using (\ref{feq}) and 
\begin{equation}
\Phi(z)\Phi(-z)=1\;,\quad \lim_{\epsilon\to 0} \frac{\Phi(\ii\eta-\ii
  n\epsilon)}{\kappa_n(\eta-\epsilon)}=1\;,\quad  
\overline{\iW}_0(\boldsymbol{x},\boldsymbol{z})\;=\;\frac{1}{n!}
  \sum_{\hat\sigma}
  \delta(\boldsymbol{x},\hat{\sigma}(\boldsymbol{z}))\;, 
\end{equation}
one obtains the inversion relation (\ref{inversion}).

\app{Series expansions of in powers of $\pp$ and $\qq$}\label{appB}
\noindent
Define power sums of exponents of the spin
variables, entering the star weight \eqref{V1},  
\begin{equation}
{\mathcal A}_k=\sum_{j=1}^n\EXP^{2\ii k a_j},\quad {\mathcal B}_k=\sum_j
\EXP^{2\ii k  b_j},\ \ldots,\quad {\mathcal X}_k=\sum_{j=1}^n\EXP^{2\ii k
  x_j},\quad 
k=1,2,\ldots\,.
\end{equation}
Consider the physical regime \eqref{regimes}, \eqref{pq-rel}. 
Then rapidity variables can be parametrized as 
\beq
\begin{array}{rclrcl}
u&=&\ds\frac{\eta}{2}+\alpha+\frac{\ii}{2}(\gamma-\beta),\qquad&
v&=&\ds-\frac{\ii}{2}(\gamma+\beta),\\[5mm]
u'&=&\ds\frac{\eta}{2}+\alpha-\frac{\ii}{2}(\gamma-\beta),&
v'&=&\ds+\frac{\ii}{2}(\gamma+\beta)
\end{array}
\eeq
where the parameters s $\alpha$, $\beta$ and $\gamma$ are real.
Moreover, assume 
\beq
|\rm{Re}{\alpha}|\ll \eta,\qquad \pp\sim\qq\to0\,, 
\eeq
Using the product expression \eqref{Phi-def} for the elliptic gamma-function 
it is not difficult to obtain the following expansions for the
Boltzmann weights \eqref{weights} and \eqref{S-def}, 
\beq
\begin{array}{l}
\kappa_n(\eta/2+\alpha)\,
\iW_{\eta/2+\alpha}(\boldsymbol{a},\boldsymbol{b})\;=\;
1+\frac{1}{2}\,(\pp\qq)^{\hf}\,\EXP^{2\alpha}\, {\mathcal A}_1^*\,{\mathcal
  B}_1^{\phantom{*}}+\pp\qq\,\EXP^{4\alpha}\, \Big\{
({\mathcal A}_1^*)^2\,({\mathcal   B}_1^{\phantom{*}})^2+
{\mathcal A}_2^*\,{\mathcal B}_2^{\phantom{*}}\Big\}\\[.5cm]
\phantom{1}+\frac{1}{6}\,(\pp\qq)^{\frac{3}{2}}\, \EXP^{6\alpha}\, \Big\{
({\mathcal A}_1^*)^3\,({\mathcal   B}_1^{\phantom{*}})^3+
2\,{\mathcal A}_3^*\,{\mathcal B}_3^{\phantom{*}}
+3\,{\mathcal A}_1^*\,{\mathcal A}_2^*\,{\mathcal B}_1^{\phantom{*}}
{\mathcal B}_2^{\phantom{*}}
\Big\}\\[.5cm]
\phantom{1}+(\pp\qq)^{\hf}\,(\pp^2+\qq^2)\,\EXP^{2\alpha}\,
{\mathcal A}_1^*\,{\mathcal B}_1^{\phantom{*}}-
(\pp\qq)^{\frac{3}{2}}\, \EXP^{-2\alpha}\,
{\mathcal A}_1^{\phantom{*}}\,{\mathcal B}_1^{{*}}\\[5mm]
\phantom{1}+\frac{1}{24}\,\pp^2\,\qq^2\,\EXP^{8\alpha}\,
\Big\{({\mathcal A}_1^*)^4\,({\mathcal   B}_1^{\phantom{*}})^4+
6\, ({\mathcal A}_1^*)^2\,{\mathcal A}_2^*\,({\mathcal
  B}_1^{\phantom{*}})^2{\mathcal   B}_2^{\phantom{*}}+
3\,({\mathcal A}_2^*)^2\,({\mathcal   B}_2^{\phantom{*}})^2
+8\, {\mathcal A}_1^*\,{\mathcal A}_3^*\,{\mathcal
  B}_1^{\phantom{*}}{\mathcal   B}_3^{\phantom{*}}
+6\, {\mathcal A}_4^*\,{\mathcal   B}_4^{\phantom{*}}\Big\}\\[5mm]
\phantom{1}+\pp\qq\,(\pp^2+\qq^2)\,\EXP^{4\alpha}
\,({\mathcal A}_1^*)^2\,({\mathcal
  B}_1^{\phantom{*}})^2-\pp^2\qq^2\,
{\mathcal A}_1^*\,{\mathcal A}_1^{\phantom{*}}\,{\mathcal
  B}_1^{*}\,{\mathcal   B}_1^{\phantom{*}}+{\mathcal O}(\pp^5)\,,
\end{array}
\eeq
where the order of the correction terms here and below is
 shown for $\pp\sim\qq$,
\beq
\begin{array}{rcl}
\kappa_n(\alpha)\,
\iW_{\alpha}(\boldsymbol{a},\boldsymbol{b})&=&
1+\pp\qq\,(1+\pp^2+\qq^2)\,\Big\{\EXP^{2\alpha}{\mathcal A}_1^*\,
{\mathcal   B}_1^{\phantom{*}}-
\EXP^{-2\alpha}{\mathcal A}_1^{\phantom{*}}\,
{\mathcal   B}_1^*\Big\}-
\pp^2\qq^2\,
{\mathcal A}_1^*\,{\mathcal A}_1^{\phantom{*}}\,{\mathcal
  B}_1^{*}\,{\mathcal   B}_1^{\phantom{*}}
\\[5mm]
&&\phantom{1}+\pp^2\qq^2\,\Big\{
\cosh 4\alpha\, ({\mathcal A}_1^*)^2\,({\mathcal
  B}_1^{\phantom{*}})^2 +
\sinh4 \alpha\,
{\mathcal A}_2^{\phantom{*}}\,{\mathcal   B}_2^{{*}}\Big\}
+{\mathcal O}(\pp^6)\,,
\end{array}
\eeq
and 
\beq
\begin{array}{l}
\ds {\pi^{n-1}n!\,\,\iS(\boldsymbol{x})}\,{
\prod_{j<k}\big(2\sin(x_j-x_k)\big)^{-2}}
\;=\;1-(\pp^2+\qq^2)\,\Big\{{\mathcal X}_1^*{\mathcal
  X}_1^{\phantom{*}}-1\Big\}+\pp^2\qq^2\,\Big\{{\mathcal X}_1^*{\mathcal
  X}_1^{\phantom{*}}-1\Big\}^2\\[5.mm]
\phantom{{\pi^{n-1}n!\,\,\iS(\boldsymbol{x})}MMMMMM}
+\frac{1}{2}\,(\pp^4+\qq^4)\,
\Big\{4-4\,{\mathcal X}_1^*{\mathcal  X}_1^{\phantom{*}}
+({\mathcal X}_1^*)^2({\mathcal
  X}_1^{\phantom{*}})^2 -{\mathcal X}_2^*{\mathcal
  X}_2^{\phantom{*}}\Big\}+{\mathcal O}(\pp^6)\,.
\end{array}
\eeq
From \eqref{pf-edge} one obtains 
\beq
\kappa(\eta/2+\alpha)=1+(\pp\qq)^{\frac{n}{2}}\,\EXP^{2n\alpha}
+{\mathcal O}(\pp^{n+2})\,.
\eeq

The definition of the IRF weight \eqref{weight-IRF} requires an evaluation 
of integrals over the central spin in the star. The results depend on
the value of $n$. Below we restrict ourselves to the case $n=3$.
For any function $f(\boldsymbol{x})$ of the spin variable $\boldsymbol{x}$
define the following integral
\beq
{\mathcal J}\big[f(\boldsymbol{x})\big]=\frac{1}{\pi^{n-1}n!}\int 
\Big(\prod_{j<k} (2\sin(x_j-x_k))^2\Big)\,
f(\boldsymbol{x}) \,d\boldsymbol{x},\qquad 
d \boldsymbol{x}=dx_1\,dx_2\cdots\,dx_{n-1}\ .
\eeq
Below we will need the following integrals
\beq\begin{array}{l}
{\mathcal J}[1]={\mathcal J}\Big[{\mathcal X_1} {\mathcal
    X_1^*}\Big]=1,\\[5mm]
{\mathcal J}\,\left[\,({\mathcal X}_1)^2 ({\mathcal X}_1^*)^2\right]=
{\mathcal J}[{\mathcal X_2} {\mathcal X_2^*}]=2,
\end{array}\label{int1}\eeq
which are given for a general $n$ and specific integrals
\beq\begin{array}{l}
{\mathcal J}[({\mathcal X}_1)^3]={\mathcal J}[({\mathcal X}_1^*)^3]=1,\\[5mm]
{\mathcal J}[{\mathcal X}_1{\mathcal X}_2]= 
{\mathcal J}[{\mathcal X}_1^*{\mathcal X}_2^*]=-1,
\end{array}\label{int2}\eeq
valid for $n=3$.
Using the above expansions one obtains for the IRF weight
for $n=3$ 
\eqref{weight-IRF}
\begin{equation}
\begin{array}{l}
\Big|{\iS(
\boldsymbol{c})\iS(\boldsymbol{b})}\Big|^{-\frac{1}{2}}\,
\mathbb{V}_{\boldsymbol{u}\boldsymbol{v}}\left(\begin{array}{cc}
\boldsymbol{a} & \boldsymbol{b} \\
\boldsymbol{c} & \boldsymbol{d}\end{array}\right)\;=\;
1+ \pp\qq\,\,{\mathcal P}
+(\pp\qq)^{\frac{3}{2}}\,{\mathcal Q}\,
\\[5mm]
\phantom{MMMMMMMMM}
+\pp^2\qq^2\,\Big(1+\frac{1}{8}\,{\mathcal P}^2+\frac{1}{4}\,{\mathcal R}
-{\mathcal S}\Big)+\frac{1}{2}\,\pp\qq\,(\pp^2+\qq^2)\,{\mathcal
  P}+{\mathcal O}(\pp^5)\label{IRF-expansion}
\end{array}
\end{equation}
where we used the following notations   
\beq
\begin{array}{rcl}
{\mathcal P}&=&\left(\EXP^{2\ii\beta}\,{\mathcal A}_1
+\EXP^{-2\ii\beta}\,{\mathcal D}_1\right)
\left(\EXP^{2\ii\gamma}\,
{\mathcal B}_1^*+\EXP^{-2\ii\gamma}\,{\mathcal C}_1^*\right)+
\left(\EXP^{-2\ii\beta}\,{\mathcal A}_1^*
+\EXP^{2\ii\beta}\,{\mathcal D}_1^*\right)
\left(\EXP^{-2\ii\gamma}\,
{\mathcal B}_1+\EXP^{2\ii\gamma}\,{\mathcal C}_1\right)\\[5mm]
{\mathcal Q}&=&\EXP^{6\alpha}\left(\EXP^{-2\ii\beta}{\mathcal A}_1
{\mathcal D}_1^*+
\EXP^{2\ii\beta}{\mathcal A}_1^*
{\mathcal D}_1\right)+\EXP^{-6\alpha}\left(\EXP^{2\ii\gamma}{\mathcal B}_1
{\mathcal C}_1^*+
\EXP^{-2\ii\gamma}{\mathcal B}_1^*
{\mathcal C}_1\right)\\[5mm]
{\mathcal R}&=&\left(\EXP^{4\ii\beta}\,{\mathcal A}_2
+\EXP^{-4\ii\beta}\,{\mathcal D}_2\right)
\left(\EXP^{4\ii\gamma}\,
{\mathcal B}_2^*+\EXP^{-4\ii\gamma}\,{\mathcal C}_2^*\right)+
\left(\EXP^{-4\ii\beta}\,{\mathcal A}_2^*
+\EXP^{4\ii\beta}\,{\mathcal D}_2^*\right)
\left(\EXP^{-4\ii\gamma}\,
{\mathcal B}_2+\EXP^{4\ii\gamma}\,{\mathcal C}_2\right)\\[5mm]
{\mathcal S}&=&\left(\EXP^{2\ii\beta}\,{\mathcal A}_1
+\EXP^{-2\ii\beta}\,{\mathcal D}_1\right)
\left(\EXP^{-2\ii\beta}\,{\mathcal A}_1^*
+\EXP^{2\ii\beta}\,{\mathcal D}_1^*\right)
+\left(\EXP^{-2\ii\gamma}\,
{\mathcal B}_1+\EXP^{2\ii\gamma}\,{\mathcal C}_1\right)
\left(\EXP^{2\ii\gamma}\,
{\mathcal B}_1^*+\EXP^{-2\ii\gamma}\,{\mathcal C}_1^*\right)
\end{array}
\eeq
All additional integrals (beyond \eqref{int1} and \eqref{int2}), 
required for \eqref{IRF-expansion} in the given order, exactly vanish.   
Note all terms in \eqref{IRF-expansion} are manifestly
real and thereby confirm the validity of the star-star relation
\eqref{star-star} in pertubation theory.

\bibliography{total32,elliptic}
\bibliographystyle{utphys}

\end{document}